\documentclass[journal,draftcls,onecolumn,12pt]{IEEEtran}

\usepackage[T1]{fontenc}
\usepackage[utf8]{inputenc}
\usepackage{epsfig}
\usepackage{amsmath}
\usepackage{amssymb}
\usepackage{cite}
\usepackage{color}
\usepackage{url}
\usepackage{amsthm}
\usepackage{algorithm}
\usepackage{algpseudocode}
\usepackage{multirow}
\usepackage{subfigure}
\usepackage{tabularx}
\usepackage{array}
\usepackage{varwidth}
\usepackage{setspace}

\makeatletter
\newcommand{\multiline}[1]{%
	\begin{tabularx}{\dimexpr\linewidth-\ALG@thistlm}[t]{@{}X@{}}
		#1
	\end{tabularx}
}
\makeatother

\graphicspath{{Figure/}}
\DeclareGraphicsExtensions{.pdf}
\theoremstyle{plain}

\theoremstyle{proposition}

\theoremstyle{definition}
\newtheorem{definition}{Definition}
\theoremstyle{remark}
\newtheorem{remark}{Remark}
 
\def\exp{{\mathrm{exp}}}

\def\log{{\mathrm{log}}}

\def\inv{^{-1}}
\def\max{{\mathrm{max}}}
\def\min{{\mathrm{min}}}
\def\tanh{{\mathrm{tanh}}}

\def\tran{^{\mbox{\scriptsize T}}}
\def\trankt{^{k,(t)\,\mbox{\scriptsize T}}}
\def\trant{^{(t)\,\mbox{\scriptsize T}}}

\def\argmin{{\mathrm{argmin}}}

\def\CE{{\mathrm{CE}}}
\def\CEDecNet{{\mathrm{CE}\text{-}\mathrm{DecNet}}}
\def\CESQLI{{\mathrm{CE}\text{-}\mathrm{SQ}\text{-}\mathrm{L1}}}
\def\CEUSQLI{{\mathrm{CE}\text{-}\mathrm{USQ}\text{-}\mathrm{L1}}}
\def\CEVQLI{{\mathrm{CE}\text{-}\mathrm{VQ}\text{-}\mathrm{L1}}}
\def\CEUSQOMP{{\mathrm{CE}\text{-}\mathrm{USQ}\text{-}\mathrm{OMP}}}
\def\ECVQ{{\mathrm{EC}\text{-}\mathrm{VQ}}}

\def\EncNet{{\mathrm{EncNet}}}

\def\DecNet{{\mathrm{DecNet}}}

\def\DeepQCS{{\mathrm{DeepVQCS}}}

\def\disp{\displaystyle}
\def\GammaE{\Gamma_{\Esf}}
\def\GammaD{\Gamma_{\Dsf}}
\def\init{{\mathrm{init}}}
\def\kt{{k,(t)}}
\def\OmegaE{\Omega_{\Esf}}
\def\OmegaD{\Omega_{\Dsf}}
\def\Nbatch{B}
\def\Ntest{N_{\mathrm{te}}}
\def\Ntrain{N_{\mathrm{tr}}}
\def\Nvalid{N_{\mathrm{va}}}
\def\test{{\mathrm{te}}}
\def\texts{\textstyle}
\def\train{{\mathrm{tr}}}
\def\valid{{\mathrm{va}}}
\def\Qbar{\bar{\Qsf}}
\def\QE{\Qsf_{\Esf}}
\def\QEbar{\bar{\Qsf}_{\Esf}}
\def\QD{\Qsf_{\Dsf}\inv}
\def\QDbar{\bar{\Qsf}_{\Dsf}\inv}

\def\Dtilde{{\tilde{D}}}
\def\Ibar{{\bar{I}}}

\def\Acal{{\mathcal{A}}}
\def\Bcal{{\mathcal{B}}}

\def\Dcal{{\mathcal{D}}}

\def\Gcal{{\mathcal{G}}}
\def\Ical{{\mathcal{I}}}

\def\Rcal{{\mathcal{R}}}

\def\Esf{{\mathsf{E}}}
\def\Dsf{{\mathsf{D}}}
\def\Qsf{{\mathsf{Q}}}
\def\Qsfbar{\bar{\mathsf{Q}}}

\def\Bbb{{\mathbb{B}}}

\def\Ebb{{\mathbb{E}}}

\def\Rbb{{\mathbb{R}}}

\newcommand{\veclatin}[1]{{\bf{#1}}}
\def\zerob{\veclatin{0}}
\def\oneb{\veclatin{1}}
\def\ab{{\veclatin{a}}}

\def\Ab{{\veclatin{A}}}

\def\bb{{\veclatin{b}}}

\def\cb{{\veclatin{c}}}

\def\Fb{{\veclatin{F}}}
\def\gb{{\veclatin{g}}}

\def\nb{{\veclatin{n}}}

\def\pb{{\veclatin{p}}}

\def\rb{{\veclatin{r}}}

\def\sb{{\veclatin{s}}}

\def\tb{{\veclatin{t}}}

\def\vb{{\veclatin{v}}}

\def\Wb{{\veclatin{W}}}
\def\xb{{\veclatin{x}}}
\def\xbhat{\hat{\veclatin{x}}}

\def\yb{{\veclatin{y}}}
\def\ybtilde{\tilde{\veclatin{y}}}

\def\zb{{\veclatin{z}}}

\newcommand{\vecgreek}[1]{{\boldsymbol{#1}}}

\def\deltab{{\vecgreek{\delta}}}

\def\lambdab{{\vecgreek{\lambda}}}
\def\Lambdab{{\vecgreek{\Lambda}}}

\def\xib{{\vecgreek{\xi}}}

\def\Phib{{\vecgreek{\Phi}}}

\begin{document}
	
\title{Low-Complexity Vector Quantized Compressed Sensing via Deep Neural Networks}
\author{Markus~Leinonen\IEEEauthorrefmark{1},~\IEEEmembership{Member,~IEEE,} and Marian~Codreanu\IEEEauthorrefmark{2},~\IEEEmembership{Member,~IEEE}

\thanks{\IEEEauthorrefmark{1}Centre for Wireless Communications -- Radio Technologies, University of Oulu, Finland. e-mail: markus.leinonen@oulu.fi.

\IEEEauthorrefmark{2}Department of Science and Technology, Link\"{o}ping University, Sweden. e-mail: marian.codreanu@liu.se.

Preliminary results of this work were presented in \cite{Leinonen-Codreanu-20}.

The work has been financially supported in part by Infotech Oulu, the Academy of Finland (grant 323698), and Academy of Finland 6Genesis Flagship (grant 318927). The work of M. Leinonen has also been financially supported in part by the Academy of Finland (grant 319485). M. Codreanu would like to acknowledge the support of the European Union's Horizon 2020 research and innovation programme under the Marie Sk\l{}odowska-Curie Grant Agreement No. 793402 (COMPRESS NETS).
}
}

\maketitle
\sloppy 

\begin{spacing}{1.4}
\begin{abstract}
Sparse signals, encountered in many wireless and signal acquisition applications, can be acquired via compressed sensing (CS) to reduce computations and transmissions, crucial for resource-limited devices, e.g., wireless sensors. Since the information signals are often continuous-valued, digital communication of compressive measurements requires quantization. In such a quantized compressed sensing (QCS) context, we address remote acquisition of a sparse source through vector quantized noisy compressive measurements. We propose a deep encoder-decoder architecture, consisting of an encoder deep neural network (DNN), a quantizer, and a decoder DNN, that realizes low-complexity vector quantization aiming at minimizing the mean-square error of the signal reconstruction for a given quantization rate. We devise a supervised learning method using stochastic gradient descent and backpropagation to train the system blocks. Strategies to overcome the vanishing gradient problem are proposed. Simulation results show that the proposed non-iterative DNN-based QCS method achieves higher rate-distortion performance with lower algorithm complexity as compared to standard QCS methods, conducive to delay-sensitive applications with large-scale signals.

\end{abstract}
\end{spacing}

\section{Introduction}
In a myriad of wireless applications and signal acquisition tasks, information signals are \emph{sparse}, i.e., they contain many zero-valued elements, either naturally or after a transformation \cite{Leinonen-Codreanu-Giannakis-19}. Sparse signals are encountered in, e.g., environmental monitoring \cite{Quer-etal-12}, source localization \cite{Malioutov-Cetin-Willsky-05}, spectrum sensing \cite{Bazerque-Giannakis-10}, and signal/anomaly detection  \cite{Wang-Li-Varshney-18}. Sparsity can be utilized by the joint sampling and compression paradigm, \emph{compressed sensing (CS)} \cite{Candes-Romberg-Tao-06,Donoho-06,Haupt-Nowak-06}, which enables accurate reconstruction of a sparse signal from few (random) linear measurements. Due to its simple encoding, allowed by more computationally intensive decoding \cite{Duarte-Wakin-Baron-Baraniuk-06}, CS is well-suited for communication applications with resource-limited encoder devices, e.g., low-power wireless sensors.

Early CS works considered continuous-valued signals and treated CS as a \emph{dimensionality reduction} technique. However, the information signals are often continuous-valued, and thus, digital communication/storage of compressive measurements requires quantization. Quantization resolution may range from simple 1-bit quantization \cite{Boufounos-Baraniuk-08} to multi-bit quantization \cite{Cheng-Ciuonzo-Rossi-20} which balances between the performance and complexity, crucial in, e.g., wireless sensors. Applying CS for digital transmission/storage initiated the \emph{quantized compressed sensing} (QCS) framework \cite{Goyal-Fletcher-Rangan-08}. QCS accomplishes \emph{source compression} in the information-theoretic sense; it compresses continuous-valued signals to finite representations. Due to \emph{indirect} observations of a source, the compression in QCS falls into \emph{remote source coding} \cite{Dobrushin-Tsybakov-62},\cite[Sect.~3.5]{Berger-71}.

Standard decoders designed for non-quantized CS yield inferior rate-distortion performance in QCS \cite{Goyal-Fletcher-Rangan-08}. Hence, it is of the utmost importance to (re)design the recovery methods for QCS to handle the impact of highly non-linear quantization, especially for low-rate schemes. The overarching idea of numerous QCS algorithms is to explicitly accommodate the presence of quantization in the encoder/decoder. First QCS works used \emph{scalar quantizers} (SQs\footnote{``SQ'' is used interchangeably to refer to ``scalar quantizer'' and ``scalar quantization''; the similar convention holds for ``VQ''.}) and optimized either the (quantization-aware) encoder or decoder \cite{Sun-Goyal-09,Zymnis-Boyd-Candes-10}. At the cost of increased complexity, enhanced rate-distortion performance is achieved by \emph{vector quantization (VQ)} \cite{Shirazinia-Chatterjee-Skoglund-14,Leinonen-Codreanu-Juntti-18,Leinonen-etal-18}. For empirical performance comparison of various QCS methods, see e.g., \cite{Leinonen-Codreanu-Juntti-19-infocom},\cite[Sect.~7.6]{Leinonen-Codreanu-Giannakis-19}.

The existing QCS methodology has two bottlenecks: 1) high encoding or/and decoding complexity, and 2) high decoding latency. Namely, although SQ permits simple encoding, the decoder runs a (quantization-aware) greedy/polynomial-complexity CS algorithm \cite{Shi-etal-16} that may become prohibitive for large-scale data and cause unacceptable delays in real-time applications. On the other hand, VQ yields superior rate-distortion performance -- even approaching the information-theoretic limit with the aid of entropy coding \cite{Leinonen-etal-18} -- but the encoding complexity grows infeasible. One remedy for this complexity-performance hindrance in QCS is \emph{deep learning}: realizing the CS decoder by a \emph{deep neural network} (DNN), along with a simple encoder/quantizer. The crux is that if such a \emph{non-iterative} signal reconstruction method meets the desired rate-distortion performance after trained \emph{offline}, the \emph{online} communications enjoy an extremely \emph{fast, low-complexity} encoding-decoding process. This would allow a resource-limited encoder device limitations, to compress and communicate large-scale data in a timely fashion.

Launching a fresh view on sparse signal reconstruction, the work \cite{Mousavi-Patel-Baraniuk-15} was the first to apply deep learning for (non-quantized) CS. The authors trained stacked denoising autoencoders to learn sparsity structures, nonlinear adaptive measurements, and a decoder to reconstruct sparse signals (images). Despite remarkable advances in non-quantized CS, only a few works have applied deep learning for QCS. The first end-to-end QCS design was proposed in \cite{Sun-etal-16}, where the devised method optimizes a binary measurement acquisition realized by a DNN, a compander-based non-uniform quantizer, and a DNN decoder to estimate neural spikes.

DNN designs applied for non-quantized CS do not resolve the peculiarities that the discretized measurements bring about. Direct adoption of ``non-quantized'' learning techniques in a QCS setup is inapplicable; the non-differentiable quantization induces a \emph{vanishing gradient problem} \cite{Bengio-Leonard-Courville-13,Agustsson-etal-17,Liu-Mattina-19}, precluding the use of standard \emph{stochastic gradient descent (SGD)} \cite[Ch.~5.9]{Goodfellow-Bengio-Courville-16} and \emph{backpropagation} \cite{Rumelhart-etal-86} in the DNN training. Since quantization becomes a pronounced factor in degrading the rate-distortion performance for low bit resolutions, the presence of a quantizer needs to be \emph{integrated} in the design. Putting all these into a QCS context, wherein the decoder receives the compressive measurements from the encoder only in  digital form, \textbf{the pertinent design task is to \emph{jointly} optimize the encoder and decoder to obtain accurate signal estimates under \emph{coarsely quantized} measurements -- which is the main focus of our paper}.

We address low-complexity remote acquisition of a sparse source through vector quantized noisy compressive measurements. To tackle the pertinent \emph{source compression} problem, we propose a \textbf{deep encoder-decoder architecture} for QCS, where 1) the encoder realizes low-complexity VQ of measurements by the cascade of an \emph{encoder DNN} and a (non-uniform) SQ, and 2) the decoder feeds the quantized measurements into a \emph{decoder DNN} to estimate the source. The key driver to our proposed DNN architecture is the fact the optimal compression in QCS is achieved by constructing a minimum mean square error estimate of the sparse source at the encoder and compressing the estimate with a VQ \cite{Dobrushin-Tsybakov-62,Wolf-Ziv-70,Leinonen-etal-18}. The objective is to train the proposed scheme to minimize the mean square error (MSE) of the signal reconstruction for a given measurement matrix and quantization rate. We use SGD and backpropagation to develop a practical supervised learning algorithm to train  \emph{jointly} the encoder DNN, quantizer, and decoder DNN. The main design driver is that once trained \emph{offline}, the non-iterative QCS method provides an extremely fast and low-complexity encoding-decoding stage for \emph{online} communications, conducive to delay-sensitive applications with large-scale signals. As the key technique to ameliorate the training, we adopt \emph{soft-to-hard quantization (SHQ)} \cite{Shlezinger-Eldar-19} at the encoder DNN to mitigate the vanishing gradient problem. The core idea is to adjust a \emph{continuous} SHQ mapping during the training to \emph{asymptotically} replicate the behavior of a continuous-to-discrete SQ implemented after training. Simulation results show that the proposed method obtains superior rate-distortion performance with faster algorithm running time compared to standard QCS methods.

\subsection{Contributions} 
To summarize, the main contributions of our paper are:
\begin{itemize} 
\item We propose a deep encoder-decoder architecture for QCS, consisting of the encoder DNN, quantizer, and decoder DNN for efficient compression of sparse signals.
\item The proposed encoder -- the cascade of the encoder DNN and SQ -- realizes low-complexity VQ, enhancing the rate-distortion performance. 
\item We provide a comprehensive treatment of the SGD optimization steps and develop a practical supervised learning algorithm to train the proposed method. 
\item We propose practical asymptotic quantizer and gradient approximation strategies for the SHQ stage to facilitate the SGD optimization and improve the performance.
\item Extensive numerical experiments illustrate that the proposed QCS method obtains superior rate-distortion performance with a low-complexity, fast encoding-decoding process in comparison to several conventional QCS methods relying on either SQ or VQ.
\item Owing to the use of VQ, the proposed DNN-based QCS method is empirically shown to be capable to approach the finite block length compression limits of QCS.   
\end{itemize}

To the best of our knowledge, DNN-based vector quantized CS has not been addressed earlier. Overall, our work gives a thorough view on the quantization aspects and challenges in the DNNs from the \emph{source compression} viewpoint, which is less explored so far. Hence, the paper is endeavored to open avenues for further developments in the related DNN context.

\subsection{Related Work}\label{sec:related_works}
In this section, we discuss the connections and differences of the related works to our paper.

\subsubsection{Learning-Based Non-Quantized CS Methods}
The first DNN-based non-quantized CS framework in \cite{Mousavi-Patel-Baraniuk-15} spurred a vast succession of (convolutional) neural network designs for compressive imaging, giving rise to, e.g., ``ReconNet'' \cite{Kulkarni-etal-16}, ``DeepCodec'' \cite{Mousavi-Dasarathy-Baraniuk-17}, ``DeepInverse'' \cite{Mousavi-Baraniuk-17}, ``SSDAE\_CS'' \cite{Zhang-etal-19}, and ``ADMM-CSNet'' \cite{Yang-etal-20}. For wireless neural recording, an autoencoder with a binary measurement matrix was devised in \cite{Sun-Feng-17}. Some works have enlightened the connections between the standard and learning-aided CS recovery; for example, ``Deep $\ell_0$ Encoder'' was proposed in \cite{Wang-Ling-Huang-16} for approximate $\ell_0$-minimization. An emergent framework \cite{Bora-etal-17} employs \textit{generative adversarial models} for compressive signal reconstruction. Besides the above point-to-point cases, deep learning has been applied for distributed CS in \cite{Palangiand-Ward-Deng-16}. These learning-based CS signal reconstruction techniques do not consider quantization.

\subsubsection{DNN-Based QCS Methods}
The most related work to our paper is \cite{Sun-etal-16}, where the developed ``BW-NQ-DNN'' method has the following differences. 1) The encoder in \cite{Sun-etal-16} has direct access to the source which allows to realize (and optimize) the measurement matrix by a DNN. This is inapplicable in our setup where the encoder observes the source only \emph{indirectly} through CS; herein, the (fixed) measurement matrix is dictated by a physical sub-sampling mechanism \cite{Mishali-Eldar-11}. 2) \cite{Sun-etal-16} quantizes measurements by a (non-uniform) SQ; we use VQ, the benefits of which are substantiated in Section~\ref{sec:results}. 3) \cite{Sun-etal-16} uses a \emph{straight-through estimator} \cite{Bengio-Leonard-Courville-13}; our SHQ with quantizer and gradient approximation policies is demonstrated to improve the (encoder) training and rate-distortion performance. 4) \cite{Sun-etal-16} uses a compander to realize non-uniform SQ; our encoder DNN (a non-linear transformation) surrogates the compander. 5) \cite{Sun-etal-16} considers noiseless CS; we consider a noisy setup.

Another related works include \cite{Cui-etal-18,Mahabadi-Lin-Cevher-19}. Different to our work, the DNN-based image recovery method in \cite{Cui-etal-18} 1) assumes direct access to the source, 2) passes the gradient through the quantizer via approximate rounding, and 3) employs SQ (in a block-by-block fashion). The work \cite{Mahabadi-Lin-Cevher-19} designs only the encoder by learning-based optimization of the CS measurement acquisition for a given decoder and uniform SQ.

\subsubsection{The ``DNN-Quantizer-DNN'' Architecture}
While the deep task-based quantization scheme in \cite{Shlezinger-Eldar-19,Shohat-etal-19} does not address source compression and CS, there is a connection to our encoder-decoder DNN architecture. The works \cite{Shlezinger-Eldar-19,Shohat-etal-19} devised a DNN-based multiple-input multiple-output communication \emph{receiver}, consisting of an analog DNN, a quantizer, and a digital DNN; this cascade is, to some extent, analogous to our encoder DNN, quantizer, and decoder DNN. Thus, \cite{Shlezinger-Eldar-19,Shohat-etal-19} address bit-constrained signal acquisition at a DNN-based \emph{decoder}, whereas we consider bit-constrained source compression at a resource-limited \emph{encoder}; our (CS-based) encoder undergoes the stringently bit-constrained quantization stage, imposed by, e.g., a low-resolution analog-to-digital converter (ADC) or/and rate-limited communications. To this end, we devise a deep \emph{joint encoder-decoder}  as a first attempt to apply DNN-based VQ in QCS.

\subsubsection{Soft-to-Hard Quantization}
We integrate the SHQ, proposed in \cite{Shohat-etal-19,Shlezinger-Eldar-19}, in our deep encoder-decoder scheme and provide techniques to optimize the SHQ to ameliorate the training. In DNNs, quantization has primarily been addressed for \emph{DNN quantization}, i.e., discretization of full-precision weights and biases for memory-efficient DNN implementation \cite{Han-Mao-Dally-15,Agustsson-etal-17,Hubara-etal-17,Liu-Mattina-19,Gong-etal-19}. These works have connections to our SHQ design. Similar to our SHQ, differential soft quantization in \cite{Gong-etal-19} uses a series of hyperbolic tangents. A soft-to-hard annealing technique was proposed for DNN and data compression in \cite{Agustsson-etal-17}; differently to our work, it uses a softmax-operator. Our gradual gradient approximation is akin to the ``alpha-blending'' method in \cite{Liu-Mattina-19}.

\subsubsection{Autoencoder}
Our DNN architecture resembles one special feedforward-type DNN -- an \emph{autoencoder} \cite{Boulard-Kamp-88}. An autoencoder attempts to copy its input to the output through an encoding and decoding function while undergoing an intermediate ``compression/representation'' stage \cite[Ch.~14]{Goodfellow-Bengio-Courville-16}. In light of remote observations, a CS scheme resembles a \emph{denoising autoencoder} \cite{Vincent-etal-10},\cite[Ch.~14.2.2]{Goodfellow-Bengio-Courville-16} which amounts to estimate the source from a \emph{corrupted} input. The ``${\ell_1\mbox{-AE}}$'' autoencoder proposed in \cite{Wu-etal-19} learns a linear encoder (i.e., the dimensionality reduction step) for a standard $\ell_1$-decoder. \emph{Uncertainty autoencoders} were employed in \cite{Grover-Ermon-19} to learn the measurement acquisition and recovery stages. Autoencoders have been designed for, e.g., CS reconstruction in \cite{Zhang-etal-19} and sparse support recovery in \cite{Li-etal-19}. All the above methods consider non-quantized CS. 

In a non-CS setup \cite{Theis-etal-17}, \emph{compressive autoencoders} were proposed for lossy image compression. 

We conclude the section by highlighting that since the inputs of our encoder DNN are the \emph{compressed} measurements, the designated task of our proposed DNN cascade is to copy a \emph{hidden/remote} information source to the decoder output. The main distinction to the above ``non-quantized autoencoders'' is that \textbf{our source compression task calls for optimizing \emph{finite representation} for the measurement vector, which itself has already undergone the dimensionality reduction stage of CS \emph{prior to} accessing the encoder.}

\textbf{Organization:} The paper is organized as follows. The system model and the problem definition are presented in Section~\ref{sec:system_model}. The proposed deep encoder-decoder architecture for QCS is introduced in Section~\ref{sec:DeepQCS_framework}. Optimization of the proposed method is detailed in Section~\ref{sec:DeepQCS_optimization}. Simulation results are presented in Section~\ref{sec:results}. Conclusions are drawn in Section~\ref{sec:conclusions}.

\textbf{Notations:}
Boldface capital letters ($\Ab$) denote matrices. 
Boldface small letters ($\ab$) denote vectors. 
Calligraphy letters ($\Acal$) denote sets.
$\Rbb_{+}$ denotes the set of non-negative real numbers.
$\zerob$ is a vector of all entries zero.
$\oneb$ is a vector of all entries one.
$(\cdot)\tran$ denotes the matrix transpose. 
$\odot$ denotes the Hadamard product.
$\circ$ denotes a composite function. 
$\tanh(\cdot)$ is the hyperbolic tangent ${\tanh(x)=\frac{e^{x}-e^{-x}}{e^{x}+e^{-x}}}$.
${f'(x)}$ denotes differentiation of function $f(x)$ with respect to $x$.
${\lceil\cdot\rceil}$ denotes rounding up to the nearest integer.
$\|\ab\|_0$ counts the number of non-zero entries of vector $\ab$. 
$\|\cdot\|_1$ and $\|\cdot\|_2$ denote the $\ell_1$-norm and $\ell_2$-norm.

\section{System Model and Problem Definition}\label{sec:system_model}
\begin{figure*}[t!]
\centering
\includegraphics[width=.95\textwidth,trim=0mm 206mm 0mm 0mm,clip]{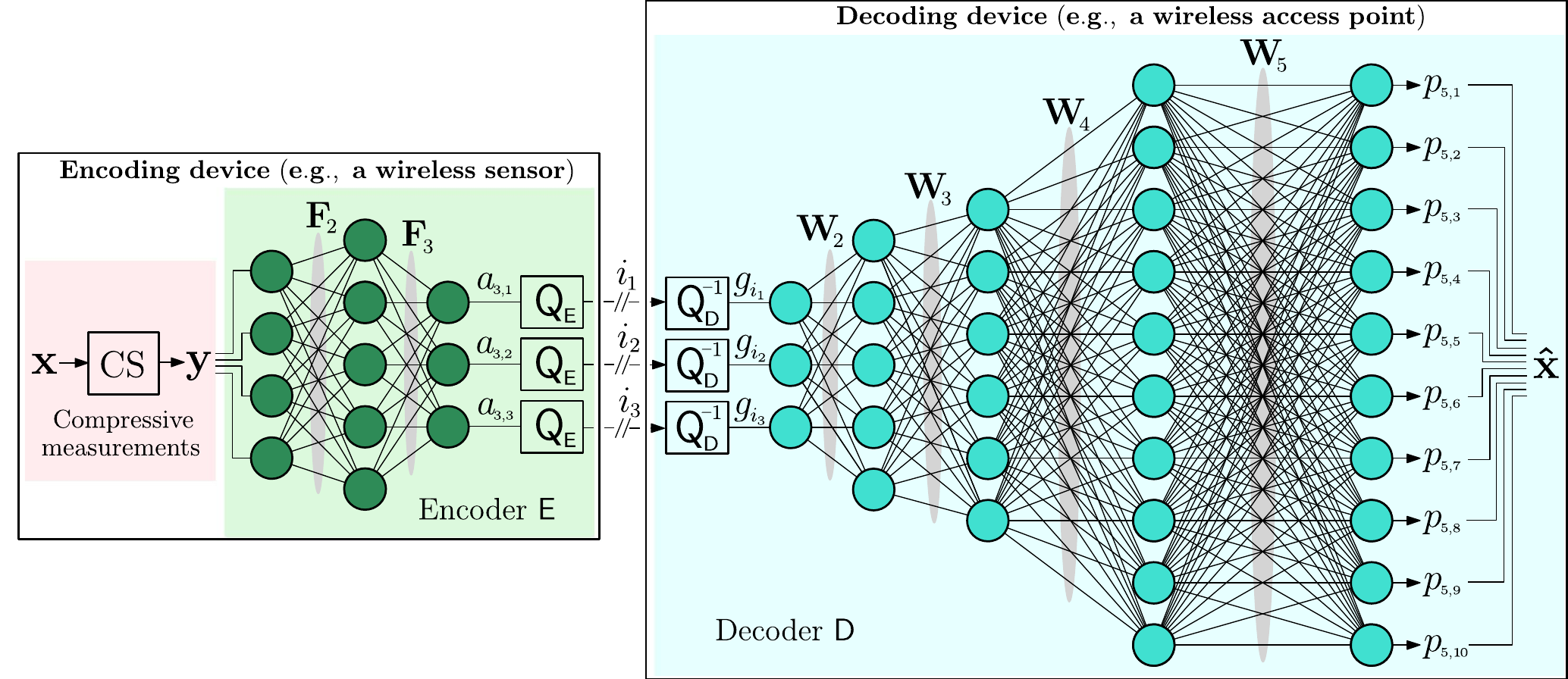}
\caption{An illustration of the proposed $\DeepQCS$ scheme for a signal of length ${N=10}$ for ${M=4}$ measurements, where 1) $\EncNet$ has ${J=3}$ layers of widths ${e_1=M=4}$, ${e_2=5}$, and ${e_3=K=3}$; and 2) $\DecNet$ has ${L=5}$ layers of widths ${d_1=K=3}$, ${d_2=5}$, ${d_3=6}$, ${d_4=10}$, and ${d_5=N=10}$.}\label{fig:System_Model}
\end{figure*}

We consider a \emph{remote} signal acquisition setup depicted in Fig.~\ref{fig:System_Model}. Encoder $\Esf$ (e.g., a low-power wireless sensor) observes an information source \emph{indirectly} in the form of noisy compressive (dimensionality-reducing) measurements. Encoder $\Esf$ processes the measurements by an \emph{encoder DNN}, \emph{quantizes} the DNN output, and communicates the quantized measurements to decoder $\Dsf$ (e.g., a wireless access point). Since our main focus is on source compression, we assume that the communication from encoder $\Esf$ to decoder $\Dsf$ is error-free. Decoder $\Dsf$ feeds the quantized measurements into a \emph{decoder DNN} to estimate the (remotely observed) source. As will be elaborated later, the method realizes (low-complexity) \emph{vector quantization} (VQ); hence, we dub the proposed deep encoder-decoder architecture for QCS as $\mathrm{\textbf{DeepVQCS}}$.

Next, we present the sensing setup and state the considered QCS problem; the DNNs and quantization stage are detailed in Section~\ref{sec:DeepQCS_framework}.

\subsection{Source Signal and Compressive Measurements}
Let ${\xb\in\Rbb^{N}}$ denote a source vector, representing, e.g., a sequence of temperature values at consecutive discrete time instants. The realizations of $\xb$ are assumed to be independent and identically distributed across time. We assume that vector\footnote{For simplicity, we assume that $\xb$ itself is sparse in a canonical basis.} $\xb$ is ${S\mbox{-sparse}}$, i.e., it has at most $S$ non-zero entries, ${{\|\xb\|}_0={S}\le{N}}$. The \emph{a priori} probabilities of the sparsity patterns are unknown.

Encoder $\Esf$ observes remote source $\xb$ (only) indirectly in the form of \emph{noisy compressive measurements} as \cite{Candes-Romberg-Tao-06,Donoho-06}
\begin{equation}\label{eq:measurements}
\yb = \Phib\xb + \nb,
\end{equation}
where ${\yb\in\Rbb^{M}}$, ${M<N}$, is a measurement vector, ${\Phib\in\Rbb^{M\times{N}}}$ is a fixed and known measurement matrix, and ${\nb\in\Rbb^{M}}$ is a noise vector. It is worth emphasizing that encoder $\Esf$ has no access to information source $\xb$: the encoder device samples and acquires $\xb$ merely via the CS, where ${\Phib}$ is dictated by a physical sub-sampling mechanism \cite{Mishali-Eldar-11}. Due to the indirect observations, the compression in a QCS setup is referred to as \emph{remote source coding} \cite{Dobrushin-Tsybakov-62},\cite[Sect.~3.5]{Berger-71}.

\subsection{Problem Definition}\label{sec:problem}
The QCS problem for optimizing the $\DeepQCS$ scheme depicted in Fig.~\ref{fig:System_Model} is given as follows.

\begin{definition}(QCS problem)\label{def:problem}
Given the compressive measurements \eqref{eq:measurements}, a fixed measurement matrix $\Phib$, and total number of quantization levels $\Ibar$ used to represent (i.e., \emph{compress}) $\yb$, the objective is to jointly optimize encoder $\Esf$ and decoder $\Dsf$ -- the encoder DNN, the quantizers, and the decoder DNN\footnote{The structure and operation of the encoder DNN, quantizers, and the decoder DNN will become explicit in Section~\ref{sec:DeepQCS_framework}.} -- for given DNN configurations to minimize the mean square error (MSE) of the signal reconstruction
\begin{equation}\label{eq:mse}
\begin{array}{ll}
\disp{D}(\Esf,\Dsf)=\Ebb\left[\left\|\Dsf\big(\Esf(\yb)\big)-\xb\right\|_{2}^{2}\right], 
\end{array}
\end{equation}
where the expectation\footnote{Throughout the paper, all expectations $\Ebb[\cdot]$ are taken with respect to the randomness of source $\xb$ and noise $\nb$.} is with respect to the randomness of source $\xb$ and noise $\nb$; $\Dsf\big(\Esf(\yb)\big)$ represents an estimate of source vector $\xb$ at the output of decoder $\Dsf$. 
\end{definition}

Our main interest in tackling the above  \emph{source compression} problem in QCS lies in devising the $\DeepQCS$ encoder-decoder architecture to enable a fast and low-complexity encoding-decoding stage, beneficial to real-time applications with large-scale sparse signals. Although the development is not tied to any particular QCS framework, our design is driven by a communication scenario, where encoder $\Esf$ is a resource-limited device (e.g., a wireless sensor) imposed by substantial limitations on total quantization/communication rate.

\section{Deep Encoder-Decoder Architecture for Quantized Compressed Sensing}\label{sec:DeepQCS_framework}
In this section, we detail the structure and operation of each block of the $\DeepQCS$ architecture. Implementation aspects of $\DeepQCS$ are also discussed.

\subsection{Encoder}
Encoder $\Esf$ comprises the encoder DNN and a quantizer encoder, described next.

\subsubsection{Encoder DNN}
As the first stage at encoder $\Esf$, the measurements in \eqref{eq:measurements} are fed into an encoder DNN\footnote{Regardless of the depth $J$, we refer to $\EncNet$ as a \emph{deep} neural network for brevity; similar convention is used for $\DecNet$.}, dubbed $\mathrm{\textbf{EncNet}}$. We consider a \emph{feedforward} DNN, or a \emph{multilayer perceptron} \cite[Ch.~6]{Goodfellow-Bengio-Courville-16}, i.e., the connections between the nodes, \emph{neurons}, form no loops. Moreover, $\EncNet$ is \emph{fully connected}\footnote{A standard fully connected feedforward DNN is considered due to its universality and suitability for supervised learning on fixed-size input vectors \cite[Sect.~11.2]{Goodfellow-Bengio-Courville-16}. Designs with more sophisticated DNN architectures are left for future work.}, i.e., each neuron at a layer is connected to all neurons at the next layer. $\EncNet$ has $J$ layers, i.e., its \emph{depth} is $J$. Next, we detail the structure and operation of $\EncNet$.

Let vector ${\cb_{j}=[c_{j,1}\cdots{c_{j,e_{j}}}]\tran\in\Rbb^{e_{j}}}$ be the \emph{weighted input} at layer $j$, where
$e_{j}$ denotes the number of neurons at layer $j$, i.e., the \emph{width} of layer $j$. 
For each \emph{hidden} layer ${j=2,\ldots,J}$, the weighted input at layer $j$ has a linear relationship to its preceding layer ${j-1}$ as
\begin{equation}\label{eq:EncNet_weighted_input}
\cb_{j} = \Fb_{j}\ab_{j-1} + \bb_{j},~j=2,\ldots,J,
\end{equation}
where ${\Fb_{j}\in\Rbb^{e_{j}\times{e_{j-1}}}}$ is the \emph{weight matrix} at layer $j$, ${\bb_{j}\in\Rbb^{e_{j}}}$ is the \emph{bias vector} at layer $j$, and vector ${\ab_{j}\in\Rbb^{e_{j}}}$ is the \emph{output} of layer $j$, defined as
\begin{equation}\label{eq:EncNet_output}
\ab_{1}=\yb,\quad\ab_{j} = \gamma_{j}(\cb_{j}) = \gamma_{j}\big( \Fb_{j}\ab_{j-1} + \bb_{j} \big),~j=2,\ldots,J,
\end{equation}
where $\gamma_{j}(\cdot)$ is an (element-wise) activation function at layer $j$; ${\ab_{1}=\yb}$ implies that the \emph{input layer} of $\EncNet$ is formed by measurement vector ${\yb}$ in \eqref{eq:measurements} (i.e., ${e_1=M}$).

Let ${K\triangleq{e_{J}}}$ denote the width of the \emph{output layer} ${j=J}$ of $\EncNet$. Accordingly, $\EncNet$ takes ${\yb\in\Rbb^{M}}$ in \eqref{eq:measurements} as its input and produces ${\ab_{J}={[a_{J,1}\cdots{a_{J,K}}]}\tran}$ of form \eqref{eq:EncNet_output} as an output. We define $\EncNet$ as a mapping  
\begin{equation}\label{eq:EncNet}
\OmegaE:\Rbb^{M}\rightarrow\Rbb^{K},\quad \OmegaE(\yb)=\ab_{J}.
\end{equation}

\subsubsection{Quantizer Encoder}
Digital communication of the $\EncNet$ outputs, i.e., the \emph{pre-processed} continuous-valued measurements ${\ab_{J}={[a_{J,1}\cdots{a_{J,K}}]}\tran}$ in \eqref{eq:EncNet}, to decoder $\Dsf$ necessitates \emph{quantization}, performed as the second stage at encoder $\Esf$. We consider a low-complexity quantization scheme where each element $a_{J,n}$, ${n=1,\ldots,K}$, is converted into a discrete representation by a \emph{scalar quantizer (SQ)}; we let the SQ to be \emph{non-uniform}, albeit a uniform SQ may be preferred in practice due to its simplicity. Accordingly, the quantization of vector ${\ab_{J}\in\Rbb^{K}}$ can be modeled\footnote{Note that portraying $K$ parallel quantizers in Fig.~\ref{fig:System_Model} is for illustration purposes; the considered SQ scheme under a single quantization rule can be implemented, e.g., by one serial scalar ADC \cite{Shlezinger-Eldar-Rodrigues-19} in a space-efficient manner.} as $K$ identical SQs (see Fig.~\ref{fig:System_Model}). A formal definition of the SQ is given next.

\begin{definition}\label{def:SQ}(SQ)
Let ${\Qsf=\{\QE,\QD\}}$ represent an ${I\text{-level}}$ SQ, consisting of quantizer encoder $\QE$ (located at encoder $\Esf$) and quantizer decoder $\QD$ (located at decoder $\Dsf$). Let ${\Rcal=\{\Rcal_{i}\}_{i\in\Ical}}$ be the set of quantization regions, where ${\Ical=\{1,\ldots,I\}}$ is the set of quantization indices. $\Rcal$ partitions real line with disjoint and exhaustive regions ${\Rcal_{i}=(t_{i-1},t_{i}]}$, where ${t_{i}\in\Rbb}$ is a \emph{threshold} with ${t_{1}\le{t_{2}}\le\cdots\le{t_{I-1}}}$; here, ${\Rcal_{1}=(-\infty,t_{1}]}$ and ${\Rcal_{I}=(t_{I-1},\infty)}$. Let ${\Gcal=\{g_{i}\}_{i\in\Ical}}$ be the set of discrete \emph{reproduction levels}, where ${g_i\in\Rbb}$ is the level associated with region $\Rcal_i$, ${i\in\Ical}$. The quantizer encoder is a mapping ${\QE:\Rbb\rightarrow\Ical}$; for $n$th $\EncNet$ output, $a_{J,n}$, it operates as 
\begin{equation}\label{eq:quantizer_enc}
\QE(a_{J,n})=i_{n}\in\Ical,~\text{if}~a_{J,n}\in\Rcal_{i_n},~n=1,\ldots,K.
\end{equation}
The quantizer decoder is a mapping ${\QD:\Ical\rightarrow\Gcal}$; for a received index ${i_{n}\in\Ical}$, it operates as 
\begin{equation}\label{eq:quantizer_dec}
\QD(i_n)=g_{i_n},~n=1,\ldots,K.
\end{equation}
In practice, quantization indices ${i\in\Ical}$\footnote{Since the SQs are identical, we will omit the neuron index ``$n$'' from quantization index ${i\in\Ical}$ whenever not explicitly needed.} are communicated from encoder $\Esf$ to decoder $\Dsf$ as \emph{binary} code words. We assume that this communication is lossless; the design of binary code words is outside of the scope of our paper. The total number of quantization levels used for compressing a measurement vector $\yb$ is ${\Ibar\triangleq{KI}}$.
\end{definition}

\begin{remark}\label{remark_quantization}
Hardware, energy, and computation restrictions of low-cost encoder devices may limit the quantization resolution, even down to one bit per sample \cite{Boufounos-Baraniuk-08}. In such QCS scenarios, the quantization error becomes a dominating factor (despite having a sufficient number of measurements $M$) in degrading the signal reconstruction accuracy. To mitigate this effect, the presence of a quantizer must be appropriately \emph{integrated} in the design. This is the main focus of our paper -- \textbf{devising a low-complexity QCS scheme to obtain high rate-distortion performance under severely bit-constrained \emph{source encoding}.}
\end{remark}

\subsubsection{Full Operation of Encoder}
Combining the operations of $\EncNet$ in \eqref{eq:EncNet} and quantizer encoder $\QE$ in \eqref{eq:quantizer_enc}, encoder $\Esf$ can be expressed as a composite function
\begin{equation}\label{eq:encoder}
\Esf=\QEbar\circ\OmegaE:\,\Rbb^{M}\rightarrow\Ical^K,
\end{equation}
where $\QEbar(\ab_{J})$ represents the aggregate quantization operation of vector $\ab_{J}$ as
\begin{equation}\label{eq:quantizer_enc_agg}
{\QEbar(\ab_{J})=\big\{\QE(a_{J,1}),\ldots,\QE(a_{J,K})\big\}=\{i_1,\ldots,i_K\}}.
\end{equation}
Finally, the full operation of encoder $\Esf$ is presented as 
\begin{equation}\label{eq:encoder_full}
\begin{array}{ll}
\!\!\!\{i_1,\ldots,i_K\}=
\QEbar\Big(\gamma_{J}\big(\Fb_{J}\big(\cdot\cdot\,\gamma_{3}\big(\Fb_{3}\big(\gamma_{2}(\Fb_{2}\yb+\bb_{2})\big)+\bb_{3}\big)\cdot\cdot\,\big)+\bb_{J}\big)\Big).
\end{array}
\end{equation}

\begin{remark}\label{remark:vq}
While each output element of $\EncNet$, $a_{J,n}$, ${n=1,\ldots,K}$, is \emph{separately} quantized by $\QE$ in \eqref{eq:quantizer_enc}, encoder $\Esf$ -- the cascade of $\EncNet$ and $\QE$ -- realizes (low-complexity) \textbf{\emph{vector quantization (VQ)}} of measurement vector $\yb$. Namely, according to \eqref{eq:encoder} and \eqref{eq:encoder_full}, encoder $\Esf$ maps $M$ input elements ${y_1,\ldots,y_M}$ \emph{jointly} into $K$ indices ${i_{n}\in\Ical}$, ${n=1,\ldots,K}$. ``Low-complexity'' refers to the fact that quantization of a measurement vector $\yb$ requires only a single forward pass through $\EncNet$ involving matrix multiplications and activation function operations, followed by a simple SQ stage. 
\end{remark}

\begin{remark}
Our proposed VQ-based DNN architecture is motivated by the fact that the optimal compression\footnote{This optimal compression strategy is known as ``estimate-and-compress'', and has been addressed under QCS in, e.g., \cite{Leinonen-etal-18,Kipnis-etal-17-isit}. Another strategy is to estimate the support of $\xb$ at the encoder and then encode the support pattern losslessly while quantizing the obtained non-zero values; this has been considered in, e.g., \cite{Goyal-Fletcher-Rangan-08,Leinonen-Codreanu-Juntti-19-infocom}. Stringently resource-limited encoding devices may necessitate to quantize the measurements directly, and run a quantization-aware CS reconstruction algorithm at the decoder; this ``compress-and-estimate'' QCS has been addressed in, e.g., \cite{Kipnis-Reeves-Eldar-18,Leinonen-Codreanu-Juntti-19-infocom}.} in QCS is achieved by constructing a minimum mean square error (MMSE) estimate of source $\xb$ at the encoder and compressing the estimate with a VQ \cite{Dobrushin-Tsybakov-62,Wolf-Ziv-70,Leinonen-etal-18}.
The capability of our DNN-based VQ to realize such optimal compression is mainly dictated by the $\EncNet$ configuration: $\EncNet$ must surrogate an exponentially complex MMSE estimation stage \cite{Elad-Yavneh-09} and a table-lookup stage. We demonstrate in Section~\ref{sec:results_DR} that $\DeepQCS$ has the ability to realize near-optimal compression. Note that due to the considered QCS architecture, the design of $\DeepQCS$ does not \emph{explicitly} reflect the sparse nature of $\xb$; rather, the sparsity is an intrinsic element in optimizing the DNN structures so as to form the optimal QCS-aware VQ.  
\end{remark}

\subsection{Decoder}
Decoder $\Dsf$ comprises a quantizer decoder and the decoder DNN, described next.

\subsubsection{Quantizer Decoder}
At the first stage of decoder $\Dsf$, $n$th quantizer decoder $\QD$ converts the received index ${i_n\in\Ical}$ into reproduction level ${g_{i_n}}$ according to \eqref{eq:quantizer_dec}, ${n=1,\ldots,K}$. The aggregate dequantization operation of index sequence ${\{i_1,\ldots,i_K\}}$ is defined as (cf. $\QEbar$ in \eqref{eq:quantizer_enc_agg})
\begin{equation}\label{eq:quantizer_dec_agg}
\QDbar(i_1,\ldots,i_K)=\big[\QD(i_{1})\cdots\QD(i_{K})\big]\tran\!
=[g_{i_1}\cdots{g_{i_K}}]\tran.
\end{equation}

Combining \eqref{eq:quantizer_enc_agg} and \eqref{eq:quantizer_dec_agg}, we denote the $K$ separate quantization-dequantization operations applied for vector $\ab_{J}$ collectively as ${\Qbar=\{\QEbar,\QDbar\}}$. Borrowing the nomenclature of DNNs, we refer to $\Qbar$ as the \emph{quantization layer}.

\subsubsection{Decoder DNN}
As the second stage of decoder $\Dsf$, the output of $\QDbar$ in \eqref{eq:quantizer_dec_agg} is fed into a feedforward fully connected ${L\text{-layer}}$ decoder DNN, dubbed $\mathrm{\textbf{DecNet}}$. The structure and operation of $\DecNet$ are as follows. Let ${\zb_{l}=[z_{l,1}\cdots{z_{l,d_{l}}}]\tran\in\Rbb^{d_{l}}}$ be the weighted input of layer $l$, where $d_{l}$ denotes the width of layer $l$. The weighted input at layer ${l=2,\ldots,L}$ is given as
\begin{equation}\label{eq:DecNet_weighted_input}
\zb_{l} = \Wb_{l}\pb_{l-1} + \rb_{l},~l=2,\ldots,L,
\end{equation}
where ${\Wb_{l}\in\Rbb^{d_{l}\times{d_{l-1}}}}$ is the weight matrix at layer $l$, ${\rb_{l}\in\Rbb^{d_{l}}}$ is the bias vector at layer $l$, and ${\pb_{l}\in\Rbb^{d_{l}}}$ is the output of layer $l$, defined as
\begin{equation}\label{eq:DecNet_output}
\pb_{1}=\gb, \quad \pb_{l} = \sigma_{l}(\zb_{l}) = \sigma_{l}\big( \Wb_{l}\pb_{l-1} + \rb_{l} \big),~l=2,\ldots,L,
\end{equation}
where $\sigma_{l}(\cdot)$ is an (element-wise) activation function at layer $l$; ${\pb_{1}=\gb}$ implies that the $\DecNet$ input is formed by reproduction levels  ${\gb=[g_{i_1}\cdots{g_{i_K}}]\tran}$ obtained via $\QDbar$ in \eqref{eq:quantizer_dec_agg}.

Owing to the estimation task, the $\DecNet$ output represents an estimate of source vector ${\xb\in\Rbb^{N}}$; thus, ${d_{L}=N}$. $\DecNet$ takes $K$ reproduction levels ${g_{i_n}}$, ${n=1,\ldots,K}$, in \eqref{eq:quantizer_dec_agg} as its input and produces ${\pb_{L}\in\Rbb^{N}}$ of form \eqref{eq:DecNet_output} as an output. We define $\DecNet$ as a mapping 
\begin{equation}\label{eq:DecNet}
\OmegaD:\Rbb^{K}\rightarrow\Rbb^{N},\quad \OmegaD(\gb)=\pb_{L}.
\end{equation}

\subsubsection{Full Operation of Decoder}
Combining the operations of quantizer decoder $\QDbar$ in \eqref{eq:quantizer_dec_agg} and $\DecNet$ in \eqref{eq:DecNet}, decoder $\Dsf$ can be expressed as a composite function
\begin{equation}\label{eq:decoder}
\Dsf=\OmegaD\circ\QDbar:\,\Ical^K\rightarrow\Rbb^{N}. 
\end{equation}
Finally, the full operation of decoder $\Dsf$ is expressed as 
\begin{equation}\label{eq:decoder_full}
\begin{array}{ll}
\pb_{L}=\mathsf{\sigma}_{L}\Big(\Wb_{L}\big(\cdot\cdot\,
\sigma_{3}\big(\Wb_{3}\big(\sigma_{2}(\Wb_{2}(\QDbar(i_1,\ldots,i_K))+\rb_{2})\big)+\rb_{3}\big)\cdot\cdot\,\big)+\rb_{L}\Big). 
\end{array}
\end{equation}

\subsection{Implementation Aspects}
Several remarks regarding the implementation of the proposed $\DeepQCS$ encoder-decoder architecture are in order. As per \eqref{eq:encoder}, $\EncNet$ processes measurements $\yb$ with real-valued numbers; the same holds true for reproduction levels $\gb$ at decoder $\Dsf$ as per \eqref{eq:decoder}. These assumptions can be invoked by various design considerations, discussed next.

\subsubsection{High-Resolution ADC \& Digital $\EncNet$}
The encoder device may be equipped with a high-resolution (e.g., 16-bit or 32-bit) ADC, when our model supports a \emph{digital} implementation of $\EncNet$. Encoder $\Esf$ receives \emph{finely} discretized measurements $\yb$ from the ADC, pre-processes $\yb$ digitally -- this is \emph{approximated} by $\OmegaE$ in \eqref{eq:EncNet} -- and employs \emph{coarse} (e.g., 1--8 bits) quantization of $\ab_{J}$ at (digital) quantizer encoder $\QE$ in \eqref{eq:quantizer_enc}. Here, $\QE$ primarily employs source compression. Note that only a moderate-sized $\EncNet$ might be viable at a low-power sensor, thereby offloading the sparse signal recovery task primarily to $\DecNet$ at a more computationally capable decoding device.

\subsubsection{Analog $\EncNet$ \& Low-Resolution ADC}
For low-cost digital devices, a low-resolution ADC precludes the above option. However, owing to the recent advances in neuromorphic computing systems, $\EncNet$ can be implemented on an \emph{analog} circuit prior to a (possibly) low-resolution ADC via \emph{memristors}\footnote{Memristor-based mixed hardware–software implementations include a two-layer DNN in \cite{Li-etal-18} and a four-layer fully connected DNN in \cite{Ambrogio-etal-18}; fully hardware implementation of a five-layer convolutional DNN was constructed in \cite{Yao-etal-20}. Inference performance of memristive DNNs is similar to that of digital DNNs, yet their computational energy efficiency and throughput per unit area are orders of magnitude higher than those of the up-to-date graphical processing units (GPUs) \cite{Ambrogio-etal-18}.} \cite{Li-etal-18,Yao-etal-20}. With exceptions of hardware non-idealities, a memristor implementation of $\EncNet$ accurately complies with the encoder mapping in \eqref{eq:encoder}. In fact, the entire encoder can be implemented via a neuromorphic system: the ADC can be realized by a memristive neural network, providing also flexible training capabilities \cite{Danial-etal-18}.

As per $\DecNet$, we assume that the receiver is equipped with a high-resolution ADC so that the operations in \eqref{eq:decoder} accurately model a digital $\DecNet$. Another practicality is that digital implementation of a DNN necessitates quantizing the weights and biases\footnote{This \emph{DNN quantization} is an active research area; see e.g., \cite{Han-Mao-Dally-15,Agustsson-etal-17,Hubara-etal-17,Liu-Mattina-19,Gong-etal-19}.}. Since our main focus is to address the presence of \emph{coarse} quantization from the source compression and communication viewpoint, a specific implementation of $\EncNet$ and $\DecNet$ as well as the inaccuracies induced by digital DNN operations are outside of the main scope of this paper and left for future work.

\section{Joint Optimization of the Deep Encoder-Decoder}\label{sec:DeepQCS_optimization}
In this section, we elaborate the optimization of the $\DeepQCS$ encoder-decoder scheme via \emph{stochastic gradient descent (SGD)} \cite[Ch.~5.9]{Goodfellow-Bengio-Courville-16} and \emph{backpropagation} \cite{Rumelhart-etal-86}. The optimization problem is formulated in Section~\ref{sec:DeepQCS_problem}. The technique to overcome the non-differentiability of quantization -- \emph{soft-to-hard quantization (SHQ)} -- is detailed in Section~\ref{sec:DeepQCS_SHQ}. The SGD optimization steps are derived in Section~\ref{sec:DeepQCS_SGD}. Asymptotic quantizer and gradient approximation strategies for the SHQ to facilitate training are proposed in Section~\ref{sec:DeepQCS_anneal}. Quantizer construction is detailed in Section~\ref{sec:DeepQCS_quantizer}. A supervised training algorithm is summarized in Section~\ref{sec:DeepQCS_algorithm}.

\subsection{Problem Formulation}\label{sec:DeepQCS_problem}
We reformulate the QCS problem in Definition~\ref{def:problem} to  incorporate the defined $\DeepQCS$ system blocks. Let $\GammaE$ and $\GammaD$ be the parameter sets of $\EncNet$ and $\DecNet$, respectively, defined as
\begin{equation}\label{eq:encoder_decoder_parameter}
\GammaE=\{\Fb_{j},\bb_{j}\}_{j=2}^{J},\,\,\,\GammaD=\{\Wb_{l},\rb_{l}\}_{l=2}^{L}.
\end{equation}
The objective of minimizing MSE distortion $D$ in \eqref{eq:mse} for a given measurement matrix $\Phib$ and quantization resolution ${\Ibar=KI}$ is cast as a joint encoder-decoder optimization problem as
\begin{equation}\label{eq:problem_global}
\begin{array}{ll}
\!\big\{\GammaE^*,\GammaD^*,\tb^*,\gb^*\big\}&\!\!\!\!=
\disp\underset{\GammaE,\GammaD,\tb,\gb}{\argmin}~\disp{D}(\GammaE,\GammaD,\tb,\gb)\\
&\!\!\!\!\overset{(a)}{=}\!\disp\underset{\GammaE,\GammaD,\tb,\gb}{\argmin}\,\,\disp\Ebb\Big[\Big\|\OmegaD\Big(\QDbar\big(\QEbar\big[\OmegaE(\yb;\GammaE);\tb\big];\gb\big);\GammaD\Big)-\xb\Big\|_{2}^{2}\Big],
\end{array}
\end{equation} 
where $(a)$ follows from the encoder and decoder mappings in \eqref{eq:encoder} and \eqref{eq:decoder}, respectively,  ${\tb=[t_{1}\cdots{t_{I-1}}]\tran}$ is the vector of thresholds of $\Qsf$ and ${\gb=[g_{1}\cdots{g_{I}}]\tran}$ is the vector of reproduction levels of $\Qsf$ (see Definition~\ref{def:SQ}).

Finding the optimal parameters of $\EncNet$ and $\DecNet$ along with the optimal quantizer\footnote{In general, finding  optimal thresholds $\tb^*$ and reproduction levels $\gb^*$ of any quantizer \emph{jointly} is difficult, if not intractable. A common approach is to optimize a quantizer via alternating optimization  by the Lloyd-Max algorithm \cite{Max-60,Lloyd-82}.} in \eqref{eq:problem_global} seems \emph{intractable}.
This is due to the complicated and \emph{non-differentiable} nature of quantizer $\Qsf$. In particular, the non-differentiability of $\Qsf$ precludes the use of standard SGD to optimize the $\DeepQCS$ scheme due to the \emph{vanishing gradient problem}\footnote{Computing the gradient of a loss function with respect to the input of a \emph{hard-thresholding} neuron (e.g., a quantizer) causes the vanishing gradient problem \cite{Bengio-Leonard-Courville-13,Agustsson-etal-17,Liu-Mattina-19} for backpropagation. \emph{Estimated gradients} have been proposed to overcome the issue; the most common one is a simple \emph{straight-through estimator} \cite{Bengio-Leonard-Courville-13} which amounts to bypassing the hard-thresholding module.} \cite{Bengio-Leonard-Courville-13,Agustsson-etal-17,Liu-Mattina-19}: since the gradient of a quantization function vanishes almost everywhere, backpropagating ``training information'' to $\EncNet$ is impossible, and thus, $\EncNet$ cannot be effectively trained for the considered compression task. Next, we address how to overcome this hindrance at the quantization layer.

\subsection{Soft-to-Hard Quantization}\label{sec:DeepQCS_SHQ}
We overcome the incapability of SGD to handle the non-differentiable quantizer by \emph{soft-to-hard quantization (SHQ)} \cite{Shlezinger-Eldar-19,Gong-etal-19,Agustsson-etal-17,Liu-Mattina-19} -- a \emph{differentiable} approximation of ``hard'' quantizer $\Qsf$.  More precisely, in the \emph{offline} training phase, we remove quantization layer $\Qsfbar$ and replace it by a \emph{virtual} $\EncNet$ layer\footnote{Representing the SHQ as a virtual layer allows us to use the unified DNN terminology established in Section~\ref{sec:DeepQCS_framework}.} ${j=J+1}$ which we call the \emph{SHQ layer}. We adopt SHQ functionality at the SHQ layer to \emph{approximate} the behavior of the quantizer that will be implemented in practice. After training, the virtual SHQ layer is removed and the SQ, $\Qsf$, is constructed using the final SHQ parameters. The premise is that once we optimize the DNN parameters -- including the SHQ parameters -- in a setup without $\Qsfbar$, the obtained parameters are expected to provide similar performance after the SHQ layer is substituted by $\Qsfbar$. We emphasize that the SHQ layer is present \emph{only} during the training phase. The detailed description of the SHQ layer is given next. 

\begin{figure}[t!]
\centering
\includegraphics[width=.6\textwidth,trim=14mm 76mm 23mm 86mm,clip]{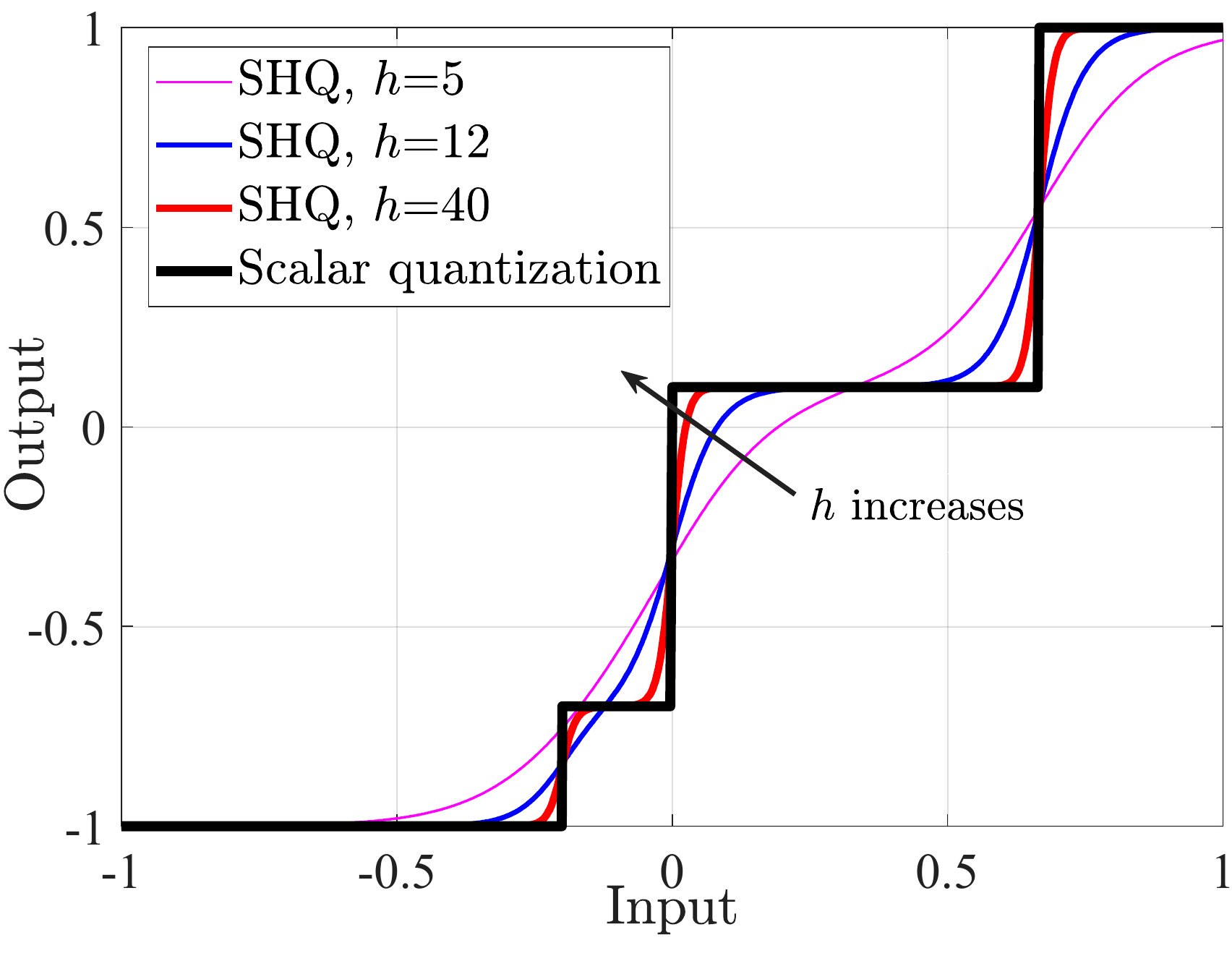}\vspace{-4mm}
\caption{Illustration of the SHQ function in \eqref{eq:SHQ} for ${I=4}$, level coefficients ${\vb=[0.15\;\;0.4\;\;0.45]\tran}$, shift coefficients ${\sb=h[-0.2\;\;0\;\;2/3]\tran}$, and steepness coefficient ${h=\{5,12,40\}}$. Adaptive adjustment of $h$ plays a key role in optimizing the $\DeepQCS$ scheme. A four-level non-uniform SQ with reproduction levels ${\{-1,-0.7,0.1,1\}}$ is depicted for comparison.}\label{fig:SHQ}
\end{figure}

Since the SHQ surrogates an SQ, the SHQ layer ${j=J+1}$ has the width ${e_{J+1}=e_{J}=K}$, and it is connected to the $\EncNet$ output layer ${j=J}$ directly, i.e., ${\cb_{J+1}=\ab_{J}}$ (cf. \eqref{eq:EncNet_weighted_input}). According to \eqref{eq:EncNet_output}, we have ${\ab_{J+1}=\gamma_{J+1}(\cb_{J+1})=\gamma_{J+1}(\ab_{J})}$, and we model the activation function $\gamma_{J+1}(\cdot)$ as the SHQ function \cite{Shlezinger-Eldar-19}; thus, $n$th SHQ output is given as
\begin{equation}\label{eq:SHQ}
\begin{array}{ll}
\!{a}_{J+1,n}\!\!\!\!\!&=\gamma_{J+1}({a}_{J,n})\\
\!\!\!\!\!&=\textstyle\sum_{i=1}^{I-1}v_{i}\tanh\big(h{a}_{J,n}-s_{i}\big),\,\,n=1,\ldots,K, 
\end{array}
\end{equation}
where \emph{level} coefficients ${\vb=[v_{1}\cdots{v_{I-1}}]\tran\in\Rbb^{I-1}_{+}}$, \emph{shift} coefficients ${\sb=[s_{1}\cdots{s_{I-1}}]\tran\in\Rbb^{I-1}}$, and \emph{steepness} coefficient ${h\in\Rbb_{+}}$ are \emph{tunable} parameters. Recall that $I$ represents the number of quantization levels of an actual SQ (see Definition~\ref{def:SQ}). The SHQ function is illustrated in Fig.~\ref{fig:SHQ}.

The SHQ function in \eqref{eq:SHQ} approximates a \emph{non-uniform} SQ as a weighted sum of shifted ${v_{i}\text{-weighted}}$ hyperbolic tangents, with the input argument being scaled by $h$. This steepness coefficient $h$ controls the \emph{asymptotic} continuous-to-discrete mapping: the higher the value of $h$, the steeper the slope of ${\tanh(h{a}_{J,n}-s_{i})}$ for a small input ${a}_{J,n}$. Thus, ${\tanh(h{a}_{J,n}-s_{i})}$ saturates quickly to $1$ ($-1$) for a small positive (negative) ${a}_{J,n}$, i.e., $\gamma_{J+1}(\cdot)$ operates like an ${I\text{-level}}$ quantizer. Level coefficients ${\vb}$ play the role of reproduction levels ${\gb=[g_{1}\cdots{g_{I}}]\tran}$ of quantizer $\Qsf$; shift coefficients ${\sb}$ are analogous to thresholds ${\tb=[t_{1}\cdots{t_{I-1}}]\tran}$ (see Definition~\ref{def:SQ}). Construction of $\Qsf$ from the SHQ parameters is detailed in Section~\ref{sec:DeepQCS_quantizer}.

\begin{remark}\label{remark:SHQ_level_shift} 
Owing to the differentiability of the SHQ function in \eqref{eq:SHQ}, one can optimize level coefficients $\vb$ and shift coefficients $\sb$ along with the other DNN parameters in \eqref{eq:encoder_decoder_parameter} within a single end-to-end SGD loop. This amounts to optimizing the quantization regions and reproduction levels of a \emph{non-uniform} SQ that will be implemented in the system.  
\end{remark}

\begin{remark}\label{remark:SHQ_steepness}
Steepness coefficient ${h}$ is a \emph{hyperparameter} and it is thus excluded from the SGD optimization. Since the vanishing gradient problem is still present at the SHQ layer (manifested in Fig.~\ref{fig:SHQ} for ${h=40}$), \emph{gradual increase} of $h$ plays a crucial role in our training procedure, as elaborated in Section~\ref{sec:DeepQCS_anneal}.
\end{remark}

\subsection{Stochastic Gradient Descent Optimization}\label{sec:DeepQCS_SGD}
In this section, we first formulate the training objective and then, use backpropagation to derive the SGD expressions needed in training the $\DeepQCS$ scheme.

\subsubsection{Training Objective}\label{sec:DeepQCS_training_cost} 
We formulate the training objective by modifying \eqref{eq:problem_global} to incorporate the SHQ layer in \eqref{eq:SHQ} while accounting for the removal of the quantizer. Let ${\Dcal_{\train}=\{\xb^{k}_{\train},\yb^{k}_{\train}\}_{k=1}^{\Ntrain}}$ be a \emph{training data set} of $\Ntrain$ source and measurement vectors sampled from their joint distribution. The \emph{training cost function} is defined as 
\begin{equation}\label{eq:cost_training}
\begin{array}{ll}
C\big(\GammaE,\GammaD,\vb,\sb\big)&\!\!\!\!=({1}/{\Ntrain})\!\sum_{k=1}^{\Ntrain}\Big\|\OmegaD\Big(\gamma_{J+1}\big(\OmegaE(\yb^{k}_{\train};\GammaE);\vb,\sb\big);\GammaD\Big)\!-\!\xb^{k}_{\train}\Big\|_2^2\\
&\!\!\!\!=({1}/{\Ntrain})\!\sum_{k=1}^{\Ntrain}\big\|\pb_{L}^{k}-\xb^{k}_{\train}\big\|_2^2,
\end{array}
\end{equation} 
where ${\pb_{L}^{k}\in\Rbb^{N}}$ is the $\DecNet$ output associated with $k$th training sample; here, we have $\OmegaD(\ab_{J+1})$ (instead of $\OmegaD(\gb)$ as per \eqref{eq:DecNet}) to account for the fact that $n$th SHQ output \eqref{eq:SHQ} is directly connected to the $n$th input of $\DecNet$, i.e., ${p_{1,n}=a_{J+1,n}}$, ${n=1,\ldots,K}$.

The training objective for the $\DeepQCS$ scheme is to find parameter sets ${\{\Gamma_{\Esf},\GammaD\}}$ in \eqref{eq:encoder_decoder_parameter} and SHQ parameters ${\{\vb,\sb\}}$ in \eqref{eq:SHQ} that minimize the training cost in \eqref{eq:cost_training} by solving the joint encoder-decoder optimization problem
\begin{equation}\label{eq:problem_training}
\big\{\GammaE^*,\GammaD^*,\vb^*,\sb^*\big\}=\underset{\GammaE,\GammaD,\vb,\sb}{\argmin}~C\big(\GammaE,\GammaD,\vb,\sb\big).
\end{equation} 
The problem \eqref{eq:problem_training} is solved by the SGD optimization, for which the required gradient updates are derived by the backpropagation. These are detailed in the next subsections.

\subsubsection{Computation of Gradients via Backpropagation}
To apply the SGD for problem \eqref{eq:problem_training}, we need to compute the gradients of cost function\footnote{To lighten the derivations, we drop the dependency of $C\big(\GammaE,\GammaD,\vb,\sb)$ on its arguments.} $C$ in \eqref{eq:cost_training} with respect to each DNN parameter set in \eqref{eq:encoder_decoder_parameter} and SHQ parameters in \eqref{eq:SHQ}. The crux of the backpropagation is to use the chain rule of the partial derivatives of $C$ to interrelate associated gradients at two consecutive layers, enabling efficient computations. Let ${\xib_{j}=[\xi_{j,1}\cdots\xi_{j,e_{j}}]\tran\in\Rbb^{e_{j}}}$ be the gradient of $C$ with respect to the weighted input of layer ${j}$ of $\EncNet$, i.e., $\cb_{j}$ in \eqref{eq:EncNet_weighted_input}, as
\begin{equation}\label{eq:grad_enc}
\begin{array}{ll}
\xib_{j}=\nabla_{\cb_{j}}C=\bigg[\disp\frac{\partial{C}}{\partial{c_{j,1}}}\cdots\frac{\partial{C}}{\partial{c_{j,e_{j}}}}\bigg]\tran,~j=2,\ldots,J+1,
\end{array}
\end{equation}
where ${\frac{\partial{C}}{\partial{c_{j,n}}}}$ denotes the partial derivative of $C$ with respect to $c_{j,n}$. Similarly, we define ${\deltab_{l}=[\delta_{l,1}\cdots\delta_{l,d_{l}}]\tran\in\Rbb^{d_{l}}}$ as the gradient of $C$ with respect to the weighted input of layer ${l}$ of $\DecNet$, i.e., $\zb_{l}$ in \eqref{eq:DecNet_weighted_input}, as 
\begin{equation}\label{eq:grad_dec}
\deltab_{l}=\nabla_{\zb_{l}}C=\bigg[\frac{\partial{C}}{\partial{z_{l,1}}}\cdots\frac{\partial{C}}{\partial{z_{l,d_{l}}}}\bigg]\tran,~l=1,\ldots,L.
\end{equation}

Next, we present the gradients in \eqref{eq:grad_enc} and \eqref{eq:grad_dec} for each layer by traversing the $\DeepQCS$ layers in the reverse order. For the $\DecNet$ output, the gradient \eqref{eq:grad_dec} is
${\deltab_{L}=\nabla_{\zb_{L}}C=2(\pb_{L}-\xb)}$. For $\DecNet$ layers ${l=1,\ldots,L-1}$, using the well-established backpropagation equations \cite{Rumelhart-etal-86}, \cite[Alg.~6.4]{Goodfellow-Bengio-Courville-16}, gradients $\deltab_{l}$ and $\deltab_{l+1}$ are interrelated as
\begin{equation}\label{eq:grad_dec_bp_vec}
\deltab_{l}=\disp\Wb_{l+1}\tran\deltab_{l+1}\odot\sigma'_{l}(\zb_{l}),~l=1,\ldots,L-1.
\end{equation}
Focus now on the SHQ layer ${j=J+1}$. Due to the absence of the quantizer, its adjacent deeper layer is the first layer of $\DecNet$, interconnected as ${\pb_{1}=\ab_{J+1}}$. Therefore, $\xib_{J+1}$ in \eqref{eq:grad_enc} is expressed as a function of $\deltab_{1}$ in \eqref{eq:grad_dec} as 
\begin{equation}\label{eq:grad_enc_bp_output_vec} 
\xib_{J+1}=\deltab_{1}\odot\gamma'_{J+1}(\ab_{J}).
\end{equation} 
Finally, for $\EncNet$ layers ${j=2,\ldots,J}$, gradients $\xib_{j}$ and $\xib_{j+1}$ are interrelated as (cf. \eqref{eq:grad_dec_bp_vec}) 
\begin{equation}\label{eq:grad_enc_bp_vec}
\xib_{j}=\disp\Fb_{j+1}\tran\xib_{j+1}\odot\gamma'_{j}(\cb_{j}),~j=2,\ldots,J.
\end{equation}

Next, we present the gradient of $C$ with respect to each DNN parameter in \eqref{eq:encoder_decoder_parameter} and SHQ parameters in \eqref{eq:SHQ} as a function of derived quantities ${\{\xib_{j}\}_{j=2}^{J+1}}$ and ${\{\deltab_{l}\}_{l=1}^{L}}$. For weight matrices and bias vectors, the gradients are given as \cite{Rumelhart-etal-86}, \cite[Alg.~6.4]{Goodfellow-Bengio-Courville-16}
\begin{equation}\label{eq:grad_enc_weight_bias}
\begin{array}{ll}
\disp\nabla_{\Fb_{j}}C=\disp\xib_{j}\ab_{j-1}\tran,~~\nabla_{\bb_{j}}C=\disp\xib_{j},~j=2,\ldots,J\\
\disp\nabla_{\Wb_{l}}C=\disp\deltab_{l}\pb_{l-1}\tran,~~\nabla_{\rb_{l}}C=\disp\deltab_{l},~l=2,\ldots,L.
\end{array}
\end{equation}

For the SHQ layer \eqref{eq:SHQ}, the gradient for level coefficients ${\nabla_{\vb}C=\big[\frac{\partial{C}}{\partial{v_{1}}}\cdots\frac{\partial{C}}{\partial{v_{I-1}}}\big]\tran}$ and the gradient for shift coefficients ${\nabla_{\sb}C=\big[\frac{\partial{C}}{\partial{s_{1}}}\cdots\frac{\partial{C}}{\partial{s_{I-1}}}\big]\tran}$ are given as follows. The partial derivative of $C$ with respect to level coefficient $v_{i}$ is derived in Appendix and is given as
\begin{equation}\label{eq:grad_SHQ_level}
\disp\frac{\partial{C}}{\partial{v_{i}}}=\texts\sum_{n=1}^{K}\delta_{1,n}\tanh\big(h{a}_{J,n}-s_{i}\big),\,\,i=1,\ldots,I-1.
\end{equation}
The partial derivative of $C$ with respect to shift coefficient  
$s_{i}$ is given as (see Appendix)
\begin{equation}\label{eq:grad_SHQ_shift}
\disp\frac{\partial{C}}{\partial{s_{i}}} 
=\texts\sum_{n=1}^{K}\delta_{1,n}v_{i}\tanh'_{s_{i}}\big(h{a}_{J,n}-s_{i}\big),\,\,i=1,\ldots,I-1,
\end{equation}
where $\tanh'_{s_{i}}(\cdot)$ denotes the derivative of $\tanh(h{a}_{J,n}-s_{i})$ with respect to $s_i$, given as $-4/{\big(\exp\{h{a}_{J,n}-s_{i}\}+\exp\{-h{a}_{J,n}+s_{i}\}\big)^2}$.

\begin{remark}\label{remark:shift_problem} 
The expression in \eqref{eq:grad_SHQ_shift} reveals that for large $h$,  $\tanh'_{s_{i}}\big(h{a}_{J,n}-s_{i}\big)$, and consequently, ${\frac{\partial{C}}{\partial{s_{i}}}}$ are close to zero almost everywhere. Thus, optimization of $\sb$ may be difficult in practice. 
\end{remark}

\subsubsection{Mini-Batch SGD Updates}
Above, we derived all required gradient expressions to apply SGD for each DNN parameter set in \eqref{eq:encoder_decoder_parameter} and SHQ parameters in \eqref{eq:SHQ} to train the $\DeepQCS$ scheme.
As a practical means, we employ the \emph{mini-batch SGD} \cite[Ch.~5.9]{Goodfellow-Bengio-Courville-16} as follows.
Let ${\Bcal^{(t)}}$ be a \emph{mini-batch} at iteration $t$, which consists of ${\Nbatch\le{\Ntrain}}$ samples $\{\xb^{k},\yb^{k}\}$ from training set $\Dcal_{\train}$. Taking weight matrix $\Fb_{j}$ as an example, the mini-batch SGD updates are of the form:  
\begin{equation}\label{eq:SGD_minibatch}
\begin{array}{ll}
\Fb_{j}^{(t+1)}&\hspace{-3mm}=\Fb_{j}^{(t)}-\Lambdab_{\Fb_{j}}^{(t)}\odot\Delta_{\Fb_{j}}^{(t)},~j=2,\ldots,J\\
&\hspace{-3mm}\overset{(a)}{=}\Fb_{j}^{(t)}-\Lambdab_{\Fb_{j}}^{(t)}\odot({1}/{\Nbatch})\sum_{k=1}^{\Nbatch}\xib_{j}^{\kt}\ab_{j-1}\trankt,
\end{array}  
\end{equation}
where superscript ${t=1,2,\ldots}$ denotes the (SGD) iteration, ${\Lambdab_{\Fb_{j}}^{(t)}\in\Rbb^{e_{j}\times{e_{j-1}}}}$ is the step size matrix at iteration $t$, ${\Delta_{\Fb_{j}}^{(t)}\in\Rbb^{e_{j}\times{e_{j-1}}}}$ is the stochastic gradient for weight matrix $\Fb_{j}$ computed over mini-batch $\Bcal^{(t)}$ at iteration $t$, $(\cdot)^{\kt}$ represents a quantity computed for $k$th sample at iteration $t$, and equality $(a)$ follows from \eqref{eq:grad_enc_weight_bias}. Thus, $\Delta_{\Fb_{j}}^{(t)}$ approximates $\nabla_{\Fb_{j}}C$ in \eqref{eq:grad_enc_weight_bias}. The mini-batch SGD updates for the other DNN parameters can be derived similarly.

\subsection{Quantizer and Gradient Approximation at the SHQ Layer}\label{sec:DeepQCS_anneal}
At the SHQ layer, steepness coefficient $h$ in \eqref{eq:SHQ} trade offs between the smoothness of the $\EncNet$-$\DecNet$ interface and the resemblance of an ${I\text{-level}}$ quantizer. Clearly, a very large value of $h$ brings the vanishing gradient problem (see Fig.~\ref{fig:SHQ} for ${h=40}$) for \eqref{eq:grad_enc_bp_output_vec}, i.e., no training information flows from $\DecNet$ to $\EncNet$, inhibiting the achievable performance. On the other hand, a small value of $h$ creates a smooth transition between the $\EncNet$ output and $\DecNet$ input (see Fig.~\ref{fig:SHQ} for ${h=5}$), passing an intact gradient flow to $\EncNet$. However, the shortcoming is that an \emph{over-relaxed} soft quantizer does not authentically represent the actual ``hard'' quantizer $\Qsf$, detrimental to the rate-distortion performance.

The aforementioned trade-off motivates to \emph{gradually increase} the presence of quantization during training. To this end, we propose two strategies that are employed to facilitate the training of $\DeepQCS$: 
1) \emph{asymptotic quantizer approximation\footnote{This is akin to \emph{annealing}, a well-established strategy in quantization; see, e.g., the VQ design in \cite{Rose-Gurewitz-Fox-92}. In DNNs, annealing has been applied, e.g., for DNN quantization in \cite{Liu-Mattina-19} and for DNN model/data compression using the softmax operator in \cite{Agustsson-etal-17}.} } that adjusts steepness coefficient $h$, and 2) \emph{gradient approximation} that (re)adjusts the gradient pass through the SHQ layer. The crux is to asymptotically increase the degree of a continuous-to-(near)-discrete mapping; initially, an ample gradient flow trains $\EncNet$ for a \emph{coarse} approximate quantizer, whereas in the course of iterations, the SHQ layer becomes an accurate replica of an ${I\text{-level}}$ quantizer and \emph{fine-tunes} $\DeepQCS$ for the given quantization resolution. These two strategies are detailed next.

\subsubsection{Asymptotic Quantizer Approximation}   
Steepness coefficient $h$ in \eqref{eq:SHQ} is updated as 
\begin{equation}\label{eq:SHQ_anneal_forward}
h^{(t)}= \min\big(h^{\init} + \alpha^{(t)},h^{\max}\big),
\end{equation}
where ${\alpha^{(t)}}$ is a step size, and parameters $h^{\init}$ and $h^{\max}$ set the initial and maximum value of $h$, respectively. A small $h^{(t)}$ approximates an identity function (see Fig.~\ref{fig:SHQ} for ${h=5}$), whereas increasing $h^{(t)}$ slowly to a large value approaches an ${I\text{-level}}$ quantizer (see Fig.~\ref{fig:SHQ} for ${h=40}$). The simulation results in Section~\ref{sec:results_gradient} show this to be an efficient strategy to ameliorate training.

\subsubsection{Gradient Approximation}  
Besides \eqref{eq:SHQ_anneal_forward}, we propose a gradual soft-to-hard transition for the backpropagating gradient through the SHQ layer. Recall that by \eqref{eq:grad_enc_bp_output_vec}, the gradient of $C$ with respect to SHQ input $\ab_{J}$ is ${\xib_{J+1}=\deltab_{1}\odot\gamma'_{J+1}(\ab_{J})}$. We propose a gradient approximation policy that uses an \emph{adjustable} weighted combination\footnote{A weighted combination for gradually increasing the impact of quantization in backpropagation is used by, e.g., the ``alpha-blending'' method \cite{Liu-Mattina-19} developed to optimize low-precision representations of a DNN model.} of the true gradient and the saturation-aware \emph{straight-through estimator (STE)}\footnote{While simple, STE has empirically been shown to be a viable means in training \cite{Bengio-Leonard-Courville-13}.} \cite{Hubara-etal-17}. Thus, at SGD iteration $t$, we have for $k$th sample:
\begin{equation}\label{eq:SHQ_anneal_backward}
\begin{array}{ll}
\xib_{J+1}^{\kt}=(1-\beta^{(t)})\big[\deltab_{1}^{\kt}\odot\oneb\big\{|\ab_{J}^{\kt}|\le{\texts\sum_{i=1}^{I-1}v_{i}}\big\}\big]+\beta^{(t)}\big[\deltab_{1}^{\kt}\odot\gamma'_{J+1}(\ab_{J}^{\kt})\big],
\end{array}
\end{equation}
where ${\beta^{(t)}=[0,1]}$ is a step size and binary vector ${\oneb\{\cdot\}\in\Bbb^{K}}$ is an element-wise indicator function: its $n$th element is zero if the magnitude of SHQ input $a_{J,n}^{\kt}$ exceeds the SHQ output range ${a_{J+1,n}\in[-\sum_{i=1}^{I-1}v_{i},\,\sum_{i=1}^{I-1}v_{i}]}$, ${n=1,\ldots,K}$ (see \eqref{eq:SQ_level_region}), i.e., it nullifies the $n$th gradient entry. For a small $\beta^{(t)}$, \eqref{eq:SHQ_anneal_backward} at early iterations tends to the STE as ${\xib_{J+1}^{\kt}\approx\deltab_{1}^{\kt}}$. One intuition at moderate values of $\beta^{(t)}$ is that the STE term of \eqref{eq:SHQ_anneal_backward} keeps passing ``noisy gradient'' for a coarse training of $\EncNet$, overriding the fact that the true gradient term of \eqref{eq:SHQ_anneal_backward} has small values around the progressively emerging flat regions of the SHQ. Finally, ${\beta^{(t)}\rightarrow1}$ ensures that the true gradient is used towards the end of training, which, along with large $h$, refines $\DecNet$ for quantization resolution $I$. In our conducted numerical experiments, combination of \eqref{eq:SHQ_anneal_forward} and \eqref{eq:SHQ_anneal_backward} with appropriate learning schedules for $\alpha^{(t)}$ and $\beta^{(t)}$ yielded the most robust training behavior.

\subsection{Quantizer Construction}\label{sec:DeepQCS_quantizer}
Once the $\DeepQCS$ scheme has been trained, the SHQ layer ${j=J+1}$ will be removed and quantizer ${\Qsf=\{\QE,\QD\}}$ is implemented in the system. Since we devoted the trainable SHQ layer to approximate an ${I\text{-level}}$ quantizer (advocated by the policies of Section~\ref{sec:DeepQCS_anneal}), we use directly the learned SHQ parameters to construct quantizer $\Qsf$ as follows.

After training, the SHQ outputs $a_{J+1,n}$ concentrate around $I$ discrete values which are dictated by level coefficients ${\{v_{i}\}_{i=1}^{I}}$ (in Fig.~\ref{fig:SHQ}, these are ${\{-1,-0.7,0.1,1\}}$); we place reproduction levels ${\{g_{i}\}_{i=1}^{I}}$ of $\Qsf$ to coincide with these saturation values. Thresholds ${\{t_{i}\}_{i=1}^{I-1}}$ of $\Qsf$ are set as ${t_{i}=s_{i}/h}$ by invoking the fact that ${\tanh\big(h{a}_{J,n}-s_{i}\big)=0}$, if ${{a}_{J,n}=s_{i}/h}$: each threshold coincides with a (nearly) vertical step occurring at an input value ${{a}_{J,n}=s_{i}/h}$ (in Fig.~\ref{fig:SHQ}, these points are ${\{-0.2,0,2/3\}}$). Formally, assuming without of loss of generality that ${{v_1}\le{v_2}\le\cdots\le{v_{I-1}}}$ and ${{s_1}\le{s_2}\le\cdots\le{s_{I-1}}}$, the quantizer $\Qsf$ is constructed as
\begin{equation}\label{eq:SQ_level_region}
\begin{array}{ll}
g_{i}=
\begin{cases}
-\sum_{i'=1}^{I-1}v_{i'},\,\,i=1\\
\sum_{i'=1}^{I-1}v_{i'} - 2\sum_{i'=i}^{I-1}v_{i'},\,\,2\le{i}\le{I{-}1}\\
\sum_{i'=1}^{I-1}v_{i'},\,\,i=I,
\end{cases} 
t_{i}=s_{i}/h,~i=1,\ldots,I-1.
\end{array}
\end{equation} 

Note that the mismatch between the SHQ and quantizer $\Qsf$ constructed according to \eqref{eq:SQ_level_region} vanishes when $h$ is sufficiently large (see Fig.~\ref{fig:SHQ}), and thus, the additional distortion incurred by implementing the actual quantizer in the system becomes minimal.

\subsection{Supervised Learning Algorithm}\label{sec:DeepQCS_algorithm}
A practical mini-batch SGD algorithm to train the $\DeepQCS$ scheme in a supervised fashion is summarized in Algorithm~\ref{alg:DeepQCS_Training}. At each iteration $t$,  training involves a \emph{forward pass} and a \emph{backward pass}, summarized in Algorithm~\ref{alg:DeepQCS_Forward} and Algorithm~\ref{alg:DeepQCS_Backward}, respectively. At iteration $t$, the rate-distortion performance of the current $\DeepQCS$ scheme with ${\{\GammaE^{(t)},\tb^{(t)},\gb^{(t)},\GammaD^{(t)}\}}$ can be evaluated using a validation set ${\Dcal_{\valid}=\{\xb^{k}_{\valid},\yb^{k}_{\valid}\}_{k=1}^{\Nvalid}}$ (or test set $\Dcal_{\test}$) as (cf. \eqref{eq:problem_global})   
\begin{equation}\label{eq:DeepQCS_DR}
\begin{array}{ll}
\Dtilde_{\valid}&\!\!\!\!=({1}/{\Nvalid})\sum_{k=1}^{\Nvalid}\big\|\pb_{L}^{\kt}-\xb^{k}_{\valid}\big\|_2^2\\
&\!\!\!\!=({1}/{\Nvalid})\sum_{k=1}^{\Nvalid}\Big\|\OmegaD\Big[\QDbar\big(\QEbar\big[\OmegaE(\yb^{k}_{\valid};\GammaE^{(t)});\tb^{(t)}\big];\gb^{(t)}\big);\GammaD^{(t)}\Big]-\xb^{k}_{\valid}\Big\|_{2}^{2}.
\end{array}
\end{equation}

The time and computation cost of the training phase can be high, typical to supervised learning. Thus, the entire training of the encoder and decoder is to be performed \emph{offline} at a computationally capable entity, e.g., a general-purpose computer. Once trained, the $\DeepQCS$ scheme communicates a measurement vector $\yb$ using only a \emph{single forward pass} in Algorithm~\ref{alg:DeepQCS_Forward} (with Step 7 replaced by quantizer $\Qsfbar$). As this involves only matrix multiplications and activation function operations, the proposed $\DeepQCS$ scheme has a fast, low-complexity encoding-decoding stage, enabling to process time-sensitive large-scale data. To assess the computational complexity and latency of the proposed method, the algorithm running time of the online phase is evaluated in Section~\ref{sec:results_time}.

\begin{algorithm}[t]\small
\caption{$\DeepQCS$ training via SGD}\label{alg:DeepQCS_Training}
\begin{algorithmic}[1]

\State \textbf{Input:} 1) Measurement matrix $\Phib$; 2) quantization levels $I$; 3) DNN configurations $J$, $L$, $K$, ${\{e_{j},\gamma_{j}\}_{j=1}^{J}}$, and ${\{d_{l},\sigma_{l}\}_{l=1}^{L}}$; 4) data sets ${\Dcal_{\train}=\{\xb^{k}_{\train},\yb^{k}_{\train}\}_{k=1}^{\Ntrain}}$, ${\Dcal_{\valid}=\{\xb^{k}_{\valid},\yb^{k}_{\valid}\}_{k=1}^{\Nvalid}}$, and ${\Dcal_{\test}=\{\xb^{k}_{\test},\yb^{k}_{\test}\}_{k=1}^{\Ntest}}$

\State Set SGD iteration index as ${t=1}$

\While{stopping criteria are not met}\Comment{{\footnotesize{\underline{Training}}}}

\State Generate a mini-batch $\Bcal^{(t)}$ from $\Dcal_{\train}$
\State Run Algorithm~\ref{alg:DeepQCS_Forward} with mini-batch $\Bcal^{(t)}$\Comment{{\footnotesize{\underline{Forward pass}}}}
\State Run Algorithm~\ref{alg:DeepQCS_Backward} with mini-batch $\Bcal^{(t)}$\Comment{{\footnotesize{\underline{Backward pass}}}}
\State \begin{varwidth}[t]{.95\linewidth} a) Construct ${\Qsf=\{\QE,\QD\}}$ in \eqref{eq:SQ_level_region}, b) run Algorithm~\ref{alg:DeepQCS_Forward} for $\Dcal_{\valid}$ by replacing Step 7 with $\Qsfbar$, and c) evaluate $\Dtilde_{\valid}$ in \eqref{eq:DeepQCS_DR}\Comment{{\footnotesize{\underline{Validation}}}}\end{varwidth}
\State Set ${t=t+1}$ 
\EndWhile 

\State Evaluate $\Dtilde_{\test}$ in \eqref{eq:DeepQCS_DR} using test set $\Dcal_{\test}$ \Comment{{\footnotesize{\underline{Testing}}}}

\State \textbf{Output:} $\DeepQCS$ encoder-decoder architecture with estimated performance $\Dtilde_{\test}$ 
\end{algorithmic}
\end{algorithm}

\begin{algorithm}[t]\small 
\caption{Forward pass at iteration $t$}\label{alg:DeepQCS_Forward}
\begin{algorithmic}[1]

\State \textbf{Input:} Mini-batch $\Bcal^{(t)}$   
\For{mini-batch sample ${k=1,\ldots,\Nbatch}$}

\State $\EncNet$ input: ${\ab_{1}^{\kt}=\yb^{k}}$  \Comment{{\footnotesize{\underline{Encoder $\Esf$}}}}
\For{$\EncNet$ layer ${j=2,\ldots,J}$}\Comment{{\footnotesize{\underline{$\EncNet$}}}}
\State $\cb_{j}^{\kt} = \Fb_{j}^{(t)}\ab_{j-1}^{\kt} + \bb_{j}^{(t)}$,\,\,\,${\ab_{j}^{\kt} = \gamma_{j}\big( \cb_{j}^{\kt} \big)}$
\EndFor

\State  ${a}_{J+1,n}^{\kt}=\sum_{i=1}^{I-1}v_{i}^{(t)}\tanh\big(h^{(t)}{a}_{J,n}^{\kt}-s_{i}^{(t)}\big),~{n=1,\ldots,K}$\Comment{{\footnotesize{\underline{SHQ layer}}}}

\State $\DecNet$ input:  ${\pb_{1}^{\kt}=\ab_{J+1}^{\kt}}$ \Comment{{\footnotesize{\underline{Decoder $\Dsf$}}}}
\For{$\DecNet$ layer ${l=2,\ldots,L}$}\Comment{{\footnotesize{\underline{$\DecNet$}}}}
\State $\zb_{l}^{\kt} = \Wb_{l}^{(t)}\pb_{l-1}^{\kt} + \rb_{l}^{(t)}$,\,\,\,$\pb_{l}^{\kt} = \sigma_{l}\big( \zb_{l}^{\kt} \big)$
\EndFor

\EndFor
\State \textbf{Output:} $\EncNet$: ${\big\{\ab_{j}^{\kt},\cb_{j}^{\kt}\big\}_{j=1}^{J}}$; SHQ: ${\ab_{J+1}^{\kt}}$;\newline
$\DecNet$: ${\big\{\pb_{l}^{\kt},\zb_{l}^{\kt}\big\}_{l=1}^{L}}$  
\end{algorithmic}
\end{algorithm} 

\begin{algorithm}[t]\small
\caption{Backward pass at iteration $t$}\label{alg:DeepQCS_Backward}
\begin{algorithmic}[1]

\State \textbf{Input:} 
1) Mini-batch $\Bcal^{(t)}$;
2) $\EncNet$: ${\big\{\ab_{j}^{\kt},\cb_{j}^{\kt}\big\}_{j=1}^{J}}$;
3) SHQ: ${\ab_{J+1}^{\kt}}$; 
4) $\DecNet$: ${\big\{\pb_{l}^{\kt},\zb_{l}^{\kt}\big\}_{l=1}^{L}}$

\For{$\DecNet$ layer ${l=L,\ldots,1}$}\Comment{{\footnotesize{\underline{Decoder $\Dsf$}}}}
\If{${l=L}$}\Comment{{\footnotesize{\underline{$\DecNet$}}}}
\State ${\deltab_{L}^{\kt}=2(\pb_{L}^{\kt}-\xb^{(t)})}$,~$\forall{k=1,\ldots,\Nbatch}$
\Else
\State $\deltab_{l}^{\kt}=\disp\Wb_{l+1}\trant\deltab_{l+1}^{\kt}\odot\sigma'_{l}\big(\zb_{l}^{\kt}\big)$,~$\forall{k=1,\ldots,\Nbatch}$
\EndIf
\If{${l>1}$}
\State \hspace{-5mm}$\Wb_{l}^{(t+1)}=\Wb_{l}^{(t)}-\Lambdab_{\Wb_{l}}^{(t)}\odot\Delta_{\Wb_{l}}^{(t)}$,\, $\rb_{l}^{(t+1)}=\rb_{l}^{(t)}-\lambdab_{\rb_{l}}^{(t)}\odot\Delta_{\rb_{l}}^{(t)}$ 
\EndIf
\EndFor

\State $\vb^{(t+1)}=\vb^{(t)}-\lambdab_{\vb}^{(t)}\odot\Delta_{\vb}^{(t)}$ \Comment{{\footnotesize{\underline{Encoder $\Esf$}}}}\newline $\sb^{(t+1)}=\sb^{(t)}-\lambdab_{\sb}^{(t)}\odot\Delta_{\sb}^{(t)}$ 
\State $\xib_{J+1}^{\kt}=(1-\beta^{(t)})\big[\deltab_{1}^{\kt}\odot\oneb\big\{|\ab_{J}^{\kt}|\le{\texts\sum_{i=1}^{I-1}v_{i}}\big\}\big]+\newline
\beta^{(t)}\big[\deltab_{1}^{\kt}\odot\gamma'_{J+1}(\ab_{J}^{\kt})\big]$,~$\forall{k=1,\ldots,\Nbatch}$\Comment{{\footnotesize{\underline{SHQ layer}}}}

\For{$\EncNet$ layer ${j=J,\ldots,2}$}\Comment{{\footnotesize{\underline{$\EncNet$}}}}

\State $\xib_{j}^{\kt}=\disp\Fb_{j+1}\trant\xib_{j+1}^{\kt}\odot\gamma'_{j}\big(\cb_{j}^{\kt}\big)$, $\forall{k=1,\ldots,\Nbatch}$
\State $\Fb_{j}^{(t+1)}=\Fb_{j}^{(t)}-\Lambdab_{\Fb_{j}}^{(t)}\odot\Delta_{\Fb_{j}}^{(t)}$, \,\,\, $\bb_{j}^{(t+1)}=\bb_{j}^{(t)}-\lambdab_{\bb_{j}}^{(t)}\odot\Delta_{\bb_{j}}^{(t)}$
\EndFor

\State \textbf{Output:} $\EncNet$: ${\big\{\Fb_{j}^{(t+1)},\bb_{j}^{(t+1)}\big\}_{j=2}^{J}}$;\newline SHQ: ${\big\{\vb^{(t+1)},\sb^{(t+1)}\big\}}$; $\DecNet$:  ${\big\{\Wb_{l}^{(t+1)},\rb_{l}^{(t+1)}\big\}_{l=2}^{L}}$ 
\end{algorithmic}
\end{algorithm}

\section{Simulation Results}\label{sec:results}
Simulation results are presented to assess the rate-distortion performance and algorithm time complexity of the proposed $\DeepQCS$ scheme summarized in Algorithm~\ref{alg:DeepQCS_Training}. The $\DeepQCS$ scheme as well as the considered baseline methods were implemented in MATLAB.

\subsection{Simulation Setup}
The simulation setup for the experiments is set as follows, unless otherwise stated.

\subsubsection{Signal Model}
For the CS setup in \eqref{eq:measurements}, we consider that 1) each non-zero entry of $\xb$ is Gaussian ${\mathcal{N}(0,1)}$, 2) the sparsity patterns are uniformly distributed, 3) each measurement noise entry is Gaussian ${\mathcal{N}(0,\sigma_{\nb}^2)}$ with ${\sigma_{\nb}^2=10^{-4}}$, and 4) $\Phib$ is generated by taking the first $M$ rows of an ${{N}\times{N}}$ discrete cosine transform matrix and normalizing the columns as ${\|\cdot\|_2^2=1}$.

\subsubsection{$\DeepQCS$}
$\EncNet$ has ${J=3}$ layers with ${e_{2}=5K}$. $\DecNet$ has ${L=5}$ layers with ${d_{2}=d_{3}=d_{4}=4N}$. The SHQ layer width is ${K=M}$. Activation functions ${\{\gamma_{j}\}_{j=2}^{J}}$ and ${\{\sigma_{l}\}_{l=2}^{L-1}}$ are $\tanh(\cdot)$; $\gamma_{1}$, $\sigma_{1}$, and $\sigma_{L}$ are identity functions. Each entry of weight matrix $\{\Fb_{j}\}_{j=2}^{J}$ ($\Wb_{l}$) is initialized by the Xavier initialization as ${\mathcal{N}(0,{1}/{e_{j-1}})}$ \cite{Glorot-Bengio-10}. The bias vectors are initialized as zero vectors. The level coefficients are initialized as ${\vb=\frac{0.8}{I-1}\oneb}$. For ${I=2}$, the shift coefficients are fixed to ${\sb=\zerob}$; for ${I>2}$, the shifts are adjusted\footnote{As pointed out in Remark~\ref{remark:shift_problem}, optimizing ${\{s_{i}\}_{i=1}^{I-1}}$ becomes challenging for large $h$. For the conducted experiments, we found that increasing $s_i$ proportional to $h^{(t)}$ to preserve the ratio ${s_{i}/h}$ and thus, to ensure well-separated SHQ regions (see Fig.~\ref{fig:SHQ} and \eqref{eq:SQ_level_region}) resulted in the best performance.}  as ${\sb^{(t)}=h^{(t)}[-0.8: \frac{1.6}{I-2} : 0.8]\tran}$. The mini-batch size is ${\Nbatch=100}$, and the data set sizes are ${\Ntrain=5\times10^{5}}$ and ${\Nvalid=\Ntest=3\times10^{5}}$. For \eqref{eq:SHQ_anneal_forward} and \eqref{eq:SHQ_anneal_backward}, we use linear step size schedules as ${\alpha^{(t)}=\alpha{t}}$ and ${\beta^{(t)}=\min\big(\beta{t},1\big)}$ with ${h^{\init}=5}$, ${\alpha=10^{-5}}$, ${h^{\max}=300}$, and ${\beta=10^{-7}}$. The step size for each DNN parameter is set by the Adam optimizer \cite[Alg.~1]{Kingma-Ba-14} and diminishing learning schedule as ($\Fb_{j}$ as an example) $\Lambdab_{\Fb_{j}}^{(t)}=\max\big\{\eta_{\Fb}^{\min},\eta_{\Fb}/{\sqrt{t}}\big\}\Acal\big(\Delta_{\Fb_{j}}^{(1)},\ldots,\Delta_{\Fb_{j}}^{(t)};\beta_{1},\beta_{2},\epsilon\big)$, where parameters $\eta_{\Fb}$ and $\eta_{\Fb}^{\min}$ adjust the initial and minimum step size, respectively; Adam $\Acal(\cdot)$ is run with the ``default'' parameters ${\beta_1=0.9}$, ${\beta_2=0.999}$, and ${\epsilon=10^{-8}}$. For weight matrices and bias vectors, we use ${\eta_{(\cdot)}=10^{-2}}$ and ${\eta_{(\cdot)}^{\min}=10^{-4}}$; for the level coefficients, we use ${\eta_{\vb}=5\times10^{-5}}$ and ${\eta_{\vb}^{\min}=5\times10^{-7}}$.

Given a signal setup ($N$, $M$, and $S$), the $\DeepQCS$ scheme is (only) empirically tuned in that the chosen learning parameters and DNN configurations (the widths, depths, activation functions etc.) remain \emph{fixed} across the quantization rates. The SGD iterations are repeated until $\Dtilde_{\valid}$ does not significantly decrease or the maximum number of iterations ${10^{7}}$ is reached.

\subsubsection{Baseline QCS Methods}
\begin{itemize} 
\item A \emph{compress-and-estimate (CE)} QCS method \cite{Leinonen-etal-18,Leinonen-Codreanu-Juntti-19-infocom} where 1) the encoder quantizes measurements $\yb$ in \eqref{eq:measurements} oblivious to $\xb$, and 2) the decoder estimates $\xb$ from quantized measurements ${\ybtilde\in\Rbb^{M}}$ through a quadratically constrained polynomial-complexity basis pursuit (BP) problem\footnote{The BP problem is solved via the ${\ell_1\text{-MAGIC}}$ package \cite{Candes-Romberg-l1magic-05} using ``l1qc\_logbarrier.m'' with stopping parameter $10^{-3}$. The problem is equivalent to the well-known basis pursuit denoising (BPDN) \cite{Chen-Donoho-Saunders-98} ${\underset{\xb\in\Rbb^N}{\mathrm{min.}}~{\mu_{\mathrm{bp}}\|\xb\|_1+\|\ybtilde-\Phib\xb\|_{2}^{2}}}$ for certain parameters $\mu_{\mathrm{qc}}$ and $\mu_{\mathrm{bp}}$ \cite[Proposition~3.2.]{Foucart-Rauhut-13}.} ${\underset{\xb\in\Rbb^N}{\mathrm{min.}}~{\|\xb\|_1}}$ s.t. ${\|\ybtilde-\Phib\xb\|_2\le{\mu_{\mathrm{qc}}}}$. Three variants are considered: 1) $\CEUSQLI$ with uniform SQ (USQ), 2) $\CESQLI$ with an SQ that is optimized to minimize the quantization distortion via the Lloyd algorithm \cite{Lloyd-82}, and 3) $\CEVQLI$ with a Lloyd-optimized VQ. We use ${\mu_{\mathrm{qc}}=\sqrt{\sigma_{\nb}}(1+1/I)}$, which is verified in Section~\ref{sec:results_baselines}.
\item A low-complexity USQ-based CE method, $\CEUSQOMP$, that estimates $\xb$ via (greedy) orthogonal matching pursuit (OMP) \cite{Pati-Rezaiifar-Krishnaprasad-93} with \emph{known sparsity $S$}.
\item A DNN-based CE method, $\CEDecNet$, where 1) the encoder uses SQ, and 2) the decoder estimates $\xb$ via the decoder DNN, $\DecNet$; we train $\CEDecNet$ similarly as $\DeepQCS$ in Algorithm~\ref{alg:DeepQCS_Training} but without $\EncNet$. This ``SQ+DNN'' architecture resembles that of ``BW-NQ-DNN'' \cite{Sun-etal-16}; however, a major difference is that ``BW-NQ-DNN'' optimizes $\Phib$, which is not applicable in our remote sensing setup. 
\item An \emph{estimate-and-compress (EC)} QCS method, $\ECVQ$ \cite{Leinonen-etal-18,Leinonen-Codreanu-Juntti-19-infocom}, where 1) the encoder forms an MMSE estimate of $\xb$ from $\yb$ which is an exponentially complex task \cite{Elad-Yavneh-09}, and 2) quantizes the resulting estimate with a Lloyd-optimized VQ. The EC strategy is known to be the \emph{optimal} compression strategy for remote source coding \cite{Dobrushin-Tsybakov-62,Wolf-Ziv-70}, while suffering from its high complexity. 
\item The \emph{remote rate-distortion function (RDF)} of  $\xb$, generated by the modified Blahut-Arimoto algorithm in \cite[Alg.~1]{Leinonen-etal-18}; this is an information-theoretic \emph{lower bound} to \emph{any} QCS method.
\end{itemize}

\subsubsection{Performance Metrics}
Reconstruction accuracy is measured as the normalized MSE (NMSE) as ${10\log_{10}\big(\Ebb[\|\xb-\xbhat\|_2^2]/\Ebb[\|\xb\|_2^2]\big)}$ (dB), where ${\xbhat\in\Rbb^{N}}$ represents a source estimate. The rate is measured as ${R=R_{\mathrm{tot}}/N}$ (bits), where $R_{\mathrm{tot}}$ is the total number of bits a QCS method uses to compress an encoder input $\yb$. For the ease of exposition, we consider that $\DeepQCS$ employs independent coding of $K$ indices $\{i_1,\ldots,i_K\}$ and thus, spends ${R_{\mathrm{tot}}=K\lceil\log_2\,I\rceil}$ bits.

\subsection{Simulation Results}

\begin{figure}[t!]
\centering
\includegraphics[width=.75\textwidth,trim=3mm 69mm 11mm 76mm,clip]{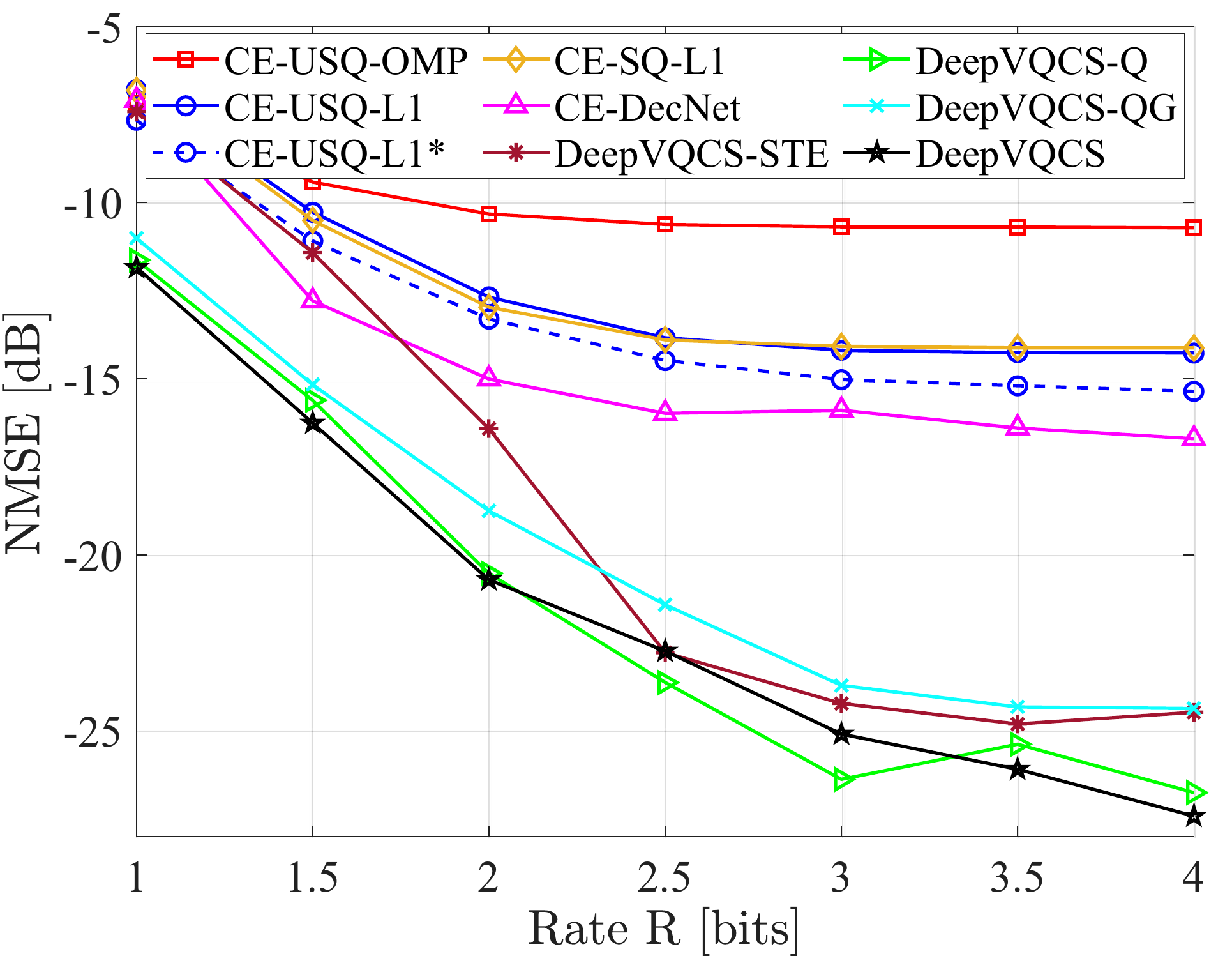}\vspace{-2mm}
\caption{Rate-distortion performance of the proposed $\DeepQCS$ method versus baseline QCS methods for ${N=20}$, ${M=10}$, and ${S=2}$. $\DeepQCS$ outperforms the considered baseline methods. The figure illustrates the importance of appropriately handling the vanishing gradient problem at the quantization layer.}\label{fig:N20_All}
\end{figure}

\subsubsection{Comparison to Baselines}\label{sec:results_baselines}
Fig.~\ref{fig:N20_All} depicts the rate-distortion performance of the $\DeepQCS$ scheme against several baseline QCS methods for ${N=20}$, ${M=10}$, and ${S=2}$. The proposed $\DeepQCS$ scheme significantly outperforms the baseline QCS methods, which are ranked in the ascending order of performance as $\CEUSQOMP$, $\CEUSQLI$, $\CESQLI$, and $\CEDecNet$. The $\CEUSQLI$ and $\CESQLI$ methods nearly coincide, indicating that SQ optimization provides negligible gain. Thus, we use $\CEUSQLI$ instead of $\CESQLI$ in sequel.

Second, we verified the choice ${\mu_{\mathrm{qc}}=\sqrt{\sigma_{\nb}}(1+1/I)}$ for $\CEUSQLI$ as follows. For each test sample $\{\xb_{\test}^{k},\yb_{\test}^{k}\}$, ${k=1,\ldots,5\times{10}^5}$, we ran the BP decoder for $56$ different values ${\mu_{\mathrm{qc}}=\{10^{-5},10^{-4.9},\ldots,10^{0.4},10^{0.5}\}}$, and read off the minimum MSE -- using the knowledge of $\xb$ -- among the candidate solutions. As Fig.~\ref{fig:N20_All} shows, this \emph{genie-aided} $\CEUSQLI^*$ variant provides only small improvement, corroborating a valid choice of $\mu_{\mathrm{qc}}$ for $\CEUSQLI$.

Third, the $\CEDecNet$ method outperforms the standard $\CE$ methods, substantiating the high potential of a DNN to replace a polynomial-complexity decoder in a QCS setup. However, because $\CEDecNet$ confines to use SQ, the gap to the proposed VQ-based $\DeepQCS$ scheme is immense: $\CEDecNet$ achieves its minimum NMSE of around ${-16.7}$ dB for ${R=4.0}$ bits, whereas $\DeepQCS$ achieves the same NMSE with more than $2.5$ times fewer bits, ${R=1.55}$. The efficacy of VQ in the $\DeepQCS$ scheme is evident in that the slope of the decay of NMSE is unrivalled; also, for the considered range of $R$, saturation is not yet encountered.

\subsubsection{Gradient Pass Strategies}\label{sec:results_gradient}
Fig.~\ref{fig:N20_All} also illustrates the impact of different gradient pass strategies at the SHQ layer for the $\DeepQCS$ scheme. Modifying \eqref{eq:SHQ_anneal_forward} and \eqref{eq:SHQ_anneal_backward}, we consider four $\DeepQCS$ variants: 
1) ${\DeepQCS\text{-}\mathrm{STE}}$ with the saturation-aware STE \cite{Hubara-etal-17} and no gradual increase of $h$ with ${h^{\init}=h^{\max}=400}$ and ${\beta=0}$;
2) ${\DeepQCS\text{-}\mathrm{Q}}$ with using only the asymptotic quantizer approximation \eqref{eq:SHQ_anneal_forward} with a modified step size schedule ${\alpha^{(t)}=0.05\,\lceil{t/100}\rceil}$ with ${h^{\init}=5}$, ${h^{\max}=300}$, and ${\beta=1}$;
3) ${\DeepQCS\text{-}\mathrm{QG}}$ with using both the quantizer and gradient approximation \eqref{eq:SHQ_anneal_forward} and \eqref{eq:SHQ_anneal_backward} with ``fast'' step size schedules ${h^{\init}=5}$, ${\alpha=10^{-4}}$, ${h^{\max}=300}$, and ${\beta=10^{-6}}$; and
4) our standard $\DeepQCS$ setting employing both \eqref{eq:SHQ_anneal_forward} and \eqref{eq:SHQ_anneal_backward} with ``slow'' step size schedules ${h^{\init}=5}$, ${\alpha=10^{-5}}$, ${h^{\max}=300}$, and ${\beta=10^{-7}}$.

\begin{figure*}[t!]
	\centering
	\subfigure[]{\includegraphics[width=.49\textwidth,trim=15mm 80mm 25mm 89mm,clip]{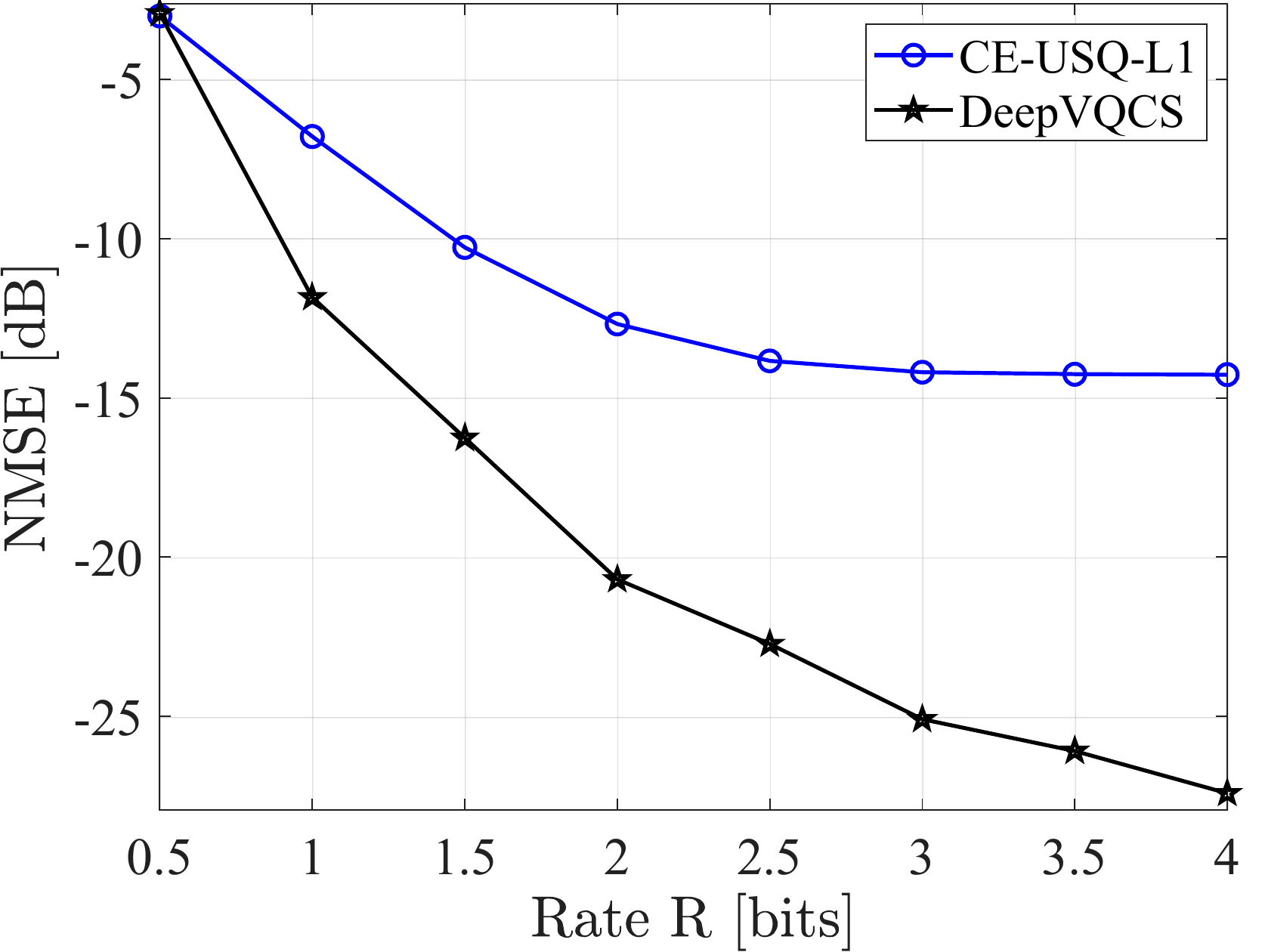}\hspace{-2mm}}
	\subfigure[]{\includegraphics[width=.49\textwidth,trim=15mm 80mm 25mm 89mm,clip]{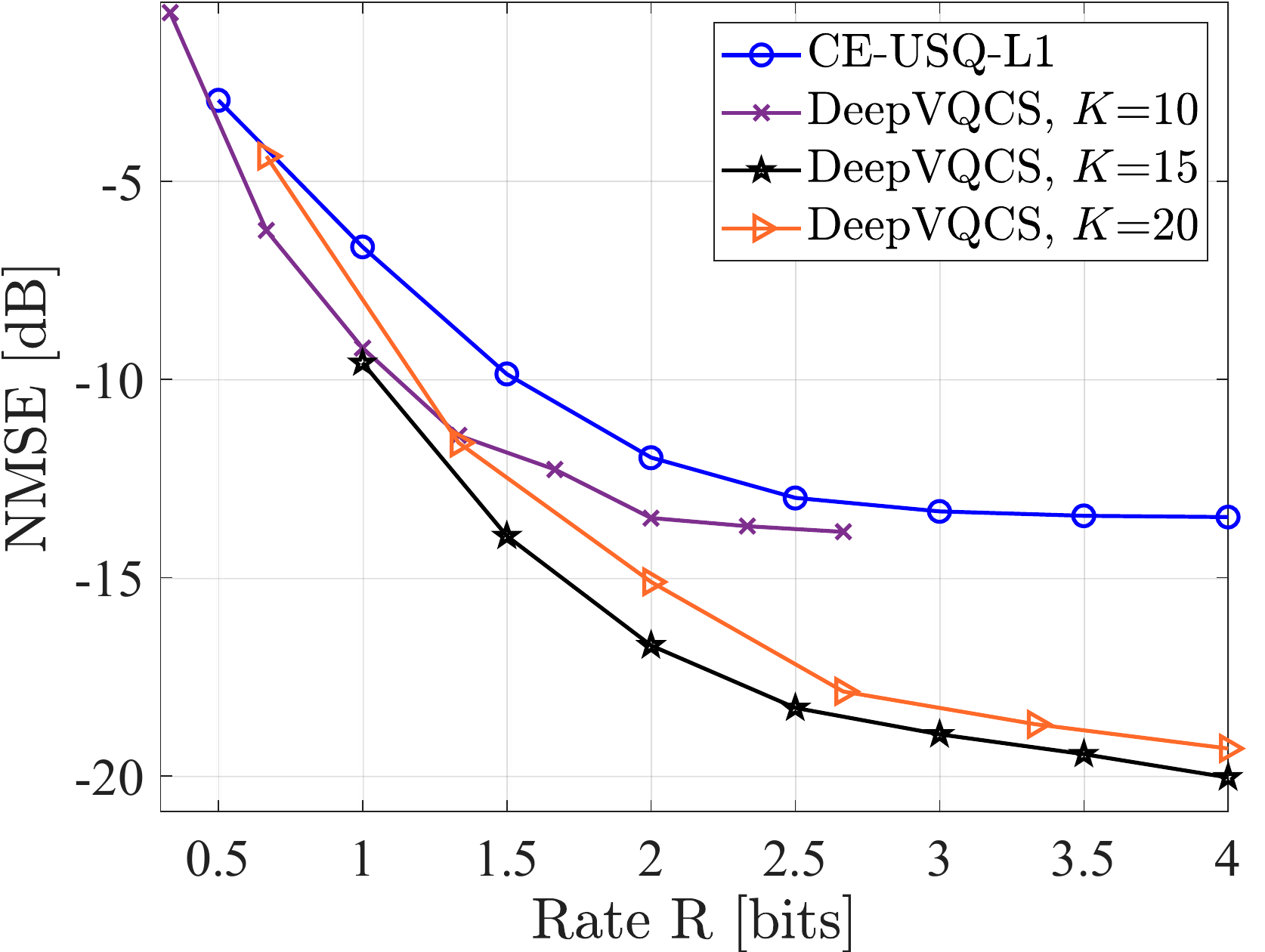}}\\\vspace{-2mm}
	\subfigure[]{\includegraphics[width=.49\textwidth,trim=15mm 80mm 25mm 89mm,clip]{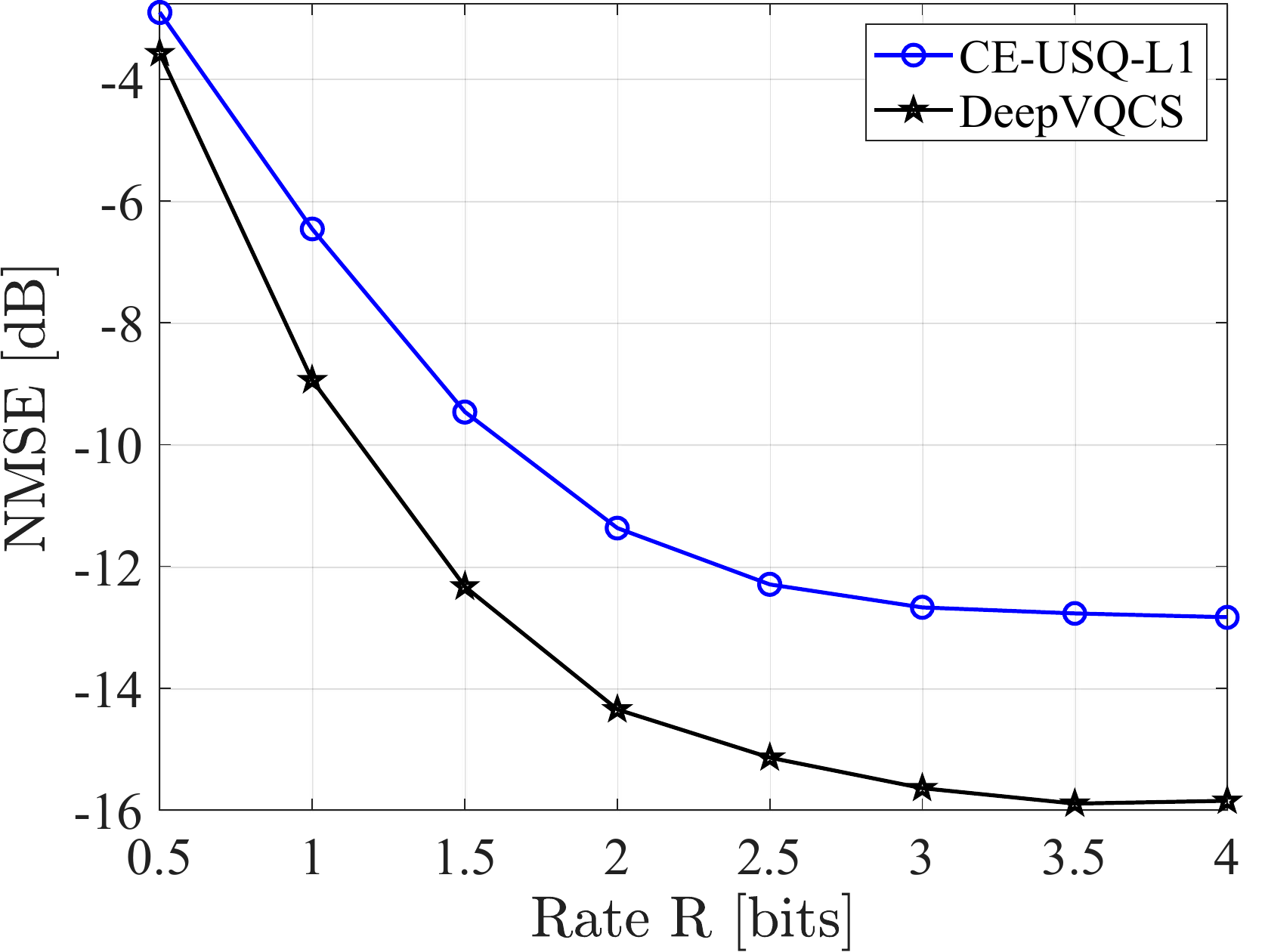}\hspace{-2mm}}
	\subfigure[]{\includegraphics[width=.49\textwidth,trim=15mm 80mm 25mm 89mm,clip]{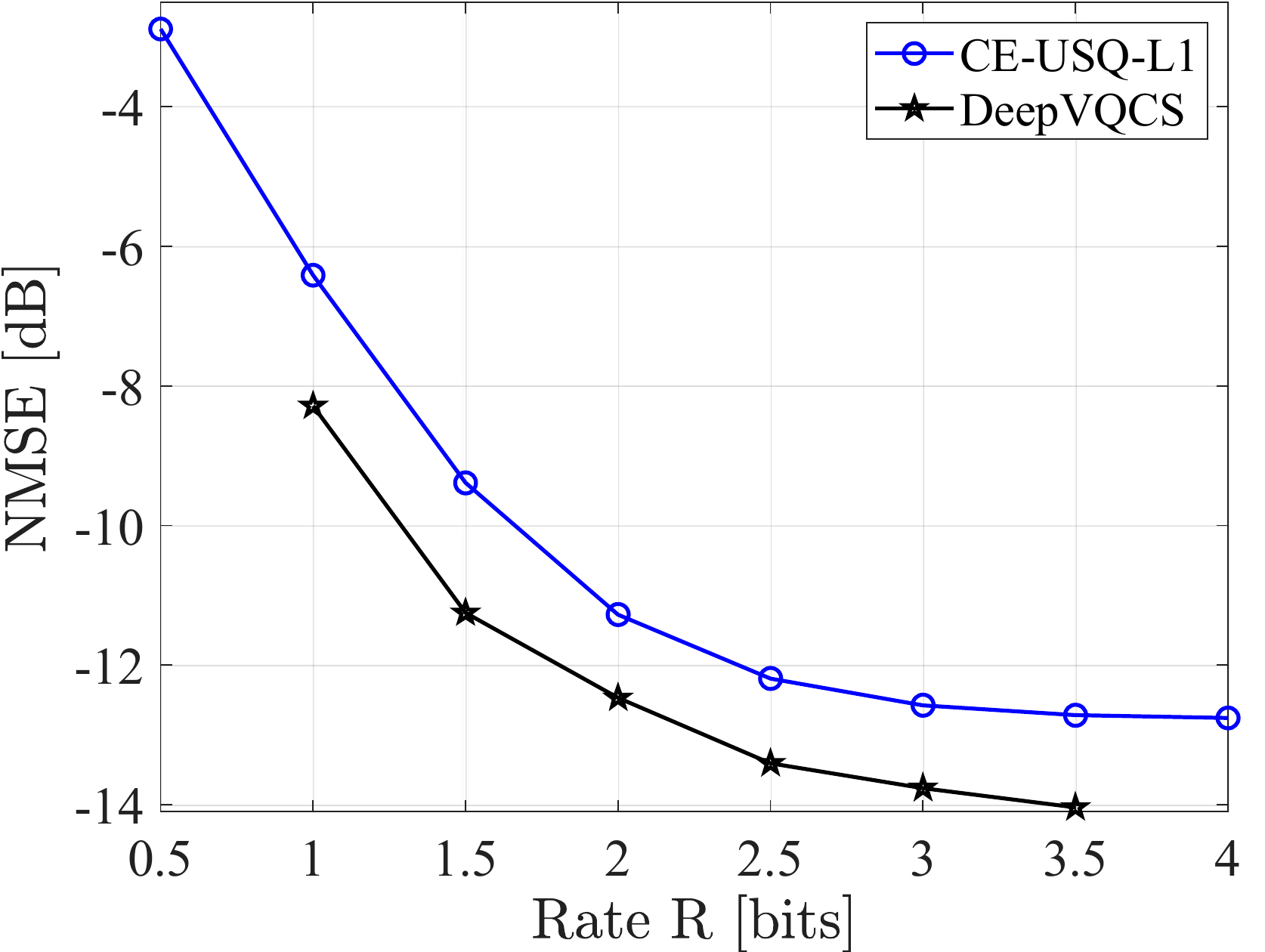}}
	\vspace{-2mm}\caption{Rate-distortion performance of the proposed $\DeepQCS$ method versus the baseline $\CEUSQLI$ method for (a) ${N=20}$, ${M=10}$, and ${S=2}$; (b) ${N=30}$, ${M=15}$, and ${S=3}$; (c) ${N=60}$, ${M=30}$, and ${S=6}$; (d) ${N=80}$, ${M=40}$, and ${S=8}$. In (b), the width of $\EncNet$ output is varied as ${K=\{10,15,20\}}$; for the other plots, we have ${K=M}$. $\DeepQCS$ scales well and outperforms the baseline in all signal setups.}\label{fig:Scaling}
\end{figure*}

Fig.~\ref{fig:N20_All} shows the benefits of the proposed strategies for the SHQ layer in \eqref{eq:SHQ_anneal_forward} and \eqref{eq:SHQ_anneal_backward} in that they provide the best rate-distortion performance. Using only the gradual increase of $h$ as per \eqref{eq:SHQ_anneal_forward} is a viable option, albeit ${\DeepQCS\text{-}\mathrm{Q}}$ encountered slightly unstable behavior at high rates. The STE cannot provide authentic training information to $\EncNet$; the shortcoming of ${\DeepQCS\text{-}\mathrm{STE}}$ is logically more pronounced for low rates. Similarly, too rapid increase of quantizer presence inhibits the performance for ${\DeepQCS\text{-}\mathrm{QG}}$. While quantitative comparison is not present, we found throughout our experiments that using the combination of \eqref{eq:SHQ_anneal_forward} and \eqref{eq:SHQ_anneal_backward} provided the most robust convergence with least sensitive choices of the learning parameters.

\subsubsection{Scalability to Different Signal Lengths}
Fig.~\ref{fig:Scaling} shows the rate-distortion performance of the $\DeepQCS$ method versus the $\CEUSQLI$ method in four different signal settings. The proposed $\DeepQCS$ method scales well to setups of different signal lengths, outperforming $\CEUSQLI$ in all setups; the gap reduces when $N$, $M$, and $S$ increase.

\subsubsection{Number of Quantizers}
Fig.~\ref{fig:Scaling}(b) shows the performance of the $\DeepQCS$ scheme for different widths of the $\EncNet$ output ${K=\{10,15,20\}}$, illustrating the trade-off for having either a few high-resolution SQs or several low-resolution SQs. The figure shows that the proposed $\DeepQCS$ scheme is flexible in terms of $K$; same rate-distortion performance can be achieved via multiple quantizer configurations. This can be beneficial for practical implementations having different hardware/operational constraints on the quantization stage.

\subsubsection{Rate-Distortion Limits}\label{sec:results_DR}
To compare the $\DeepQCS$ method against the rate-distortion limits of QCS, we consider the experiment in \cite[Fig.~5(c)]{Leinonen-etal-18} with ${N=7}$, ${M=4}$, ${S=1}$, and ${\sigma_{\nb}^2=0.01}$. We use ${J=4}$, ${e_{2}=e_{3}=5K}$, ${L=4}$, ${d_{2}=d_{3}=d_{4}=5N}$, ${\alpha=10^{-3}}$, and ${\beta=10^{-6}}$. We consider a special structure for $\EncNet$: its output layer is formed by $K$ \emph{$1$-bit SQs}. Thus, ${I=2}$ and the SHQ function in \eqref{eq:SHQ} is a single (weighted) $\tanh(\cdot)$. Note that although the signal setup is small, it allows to elucidate the fundamental compression capabilities of $\DeepQCS$.

\begin{figure}[t!]
\centering
\includegraphics[width=.75\textwidth,trim=5mm 73mm 13mm 80mm,clip]{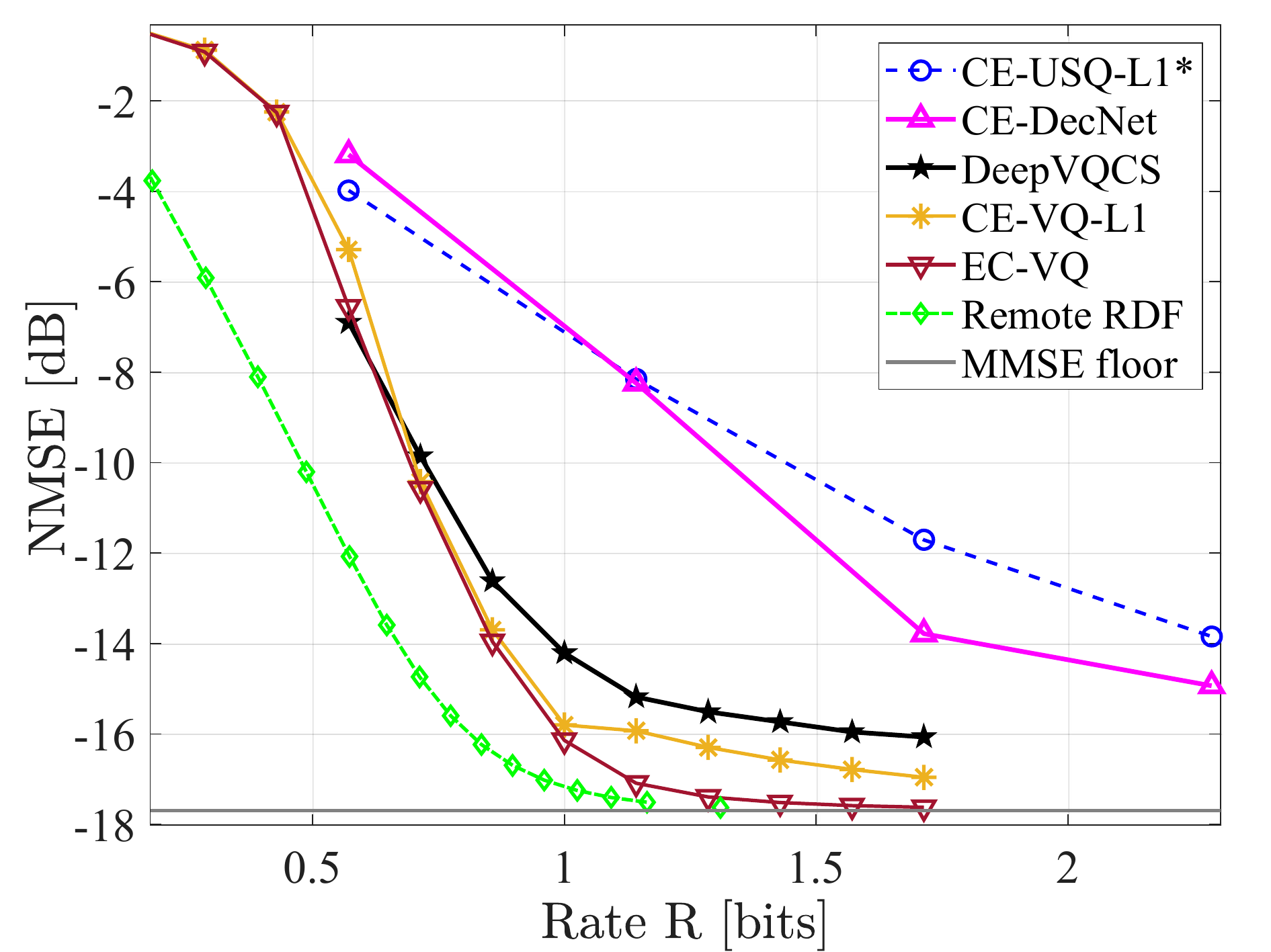}\vspace{-2mm}
\caption{Rate-distortion performance of the proposed $\DeepQCS$ method versus practical QCS methods and the information-theoretic limit of QCS (``Remote RDF'') for ${N=7}$, ${M=4}$, and ${S=1}$. The performance of $\DeepQCS$ is close to that of $\ECVQ$ which represents the optimal compression achievable in QCS by fixed-rate VQ of a measurement vector $\yb$.}\label{fig:DR}
\end{figure}

Fig.~\ref{fig:DR} shows the rate-distortion performance of the $\DeepQCS$ method against the baselines and theoretical limits of QCS. The SQ-based $\CEDecNet$ method slightly outperforms $\CEUSQLI^*$. The figure corroborates the ability of the proposed $\DeepQCS$ scheme to realize near-optimal compression: by employing VQ through the cascade of $\EncNet$ and SQ, $\DeepQCS$ takes a significant leap from $\CEDecNet$ and performs close to the tabular-search VQ-based $\CEVQLI$ and $\ECVQ$ methods. Recall that $\ECVQ$ involves exponentially complex MMSE estimation at the encoder, which becomes prohibitive for large-scale signals. In fact, the best anticipated performance for $\DeepQCS$ is to match with $\ECVQ$ as they both apply VQ of a single vector $\yb$ at a time; note that the remote RDF is portrayed as the information-theoretic limit of QCS that can be achieved only via (excessively complex) VQ of multiple vector inputs $\yb$ \cite{Leinonen-etal-18}.

\subsubsection{Algorithm Running Time}\label{sec:results_time}
To assess the computational complexity and latency for communicating a measurement vector $\yb$, Table~\ref{table:time} compares the algorithm running time\footnote{Algorithm running time was evaluated using ``tic'' function in MATLAB.} in the online phase for three different signal setups. The decoding and total running times of the proposed $\DeepQCS$ scheme are around ${35-60}$ and ${25-40}$ times lower than those of the polynomial-complexity $\CEUSQLI$ method, respectively. The $\DeepQCS$ scheme is faster than the greedy $\CEUSQOMP$ method (which runs only $S$ loops). Additional pre-processing via $\EncNet$ inevitably increases the encoding time of $\DeepQCS$. Note, however, that the encoding time of each algorithm is small in proportion to the decoding time, i.e., the decoding time dominates the total latency incurred in the encoding-decoding process.

\begin{table*}[t!]\scriptsize
	\begin{center} 
		\caption{Algorithm running time comparison in the online phase: encoding/decoding/total time of $\CEUSQOMP$ (first row) and $\CEUSQLI$ (second row) normalized with respect to those of the $\DeepQCS$ scheme}
		\vspace{-3mm}\begin{tabular}{ |    
				p{1.38cm}|p{1.38cm}|p{1.38cm}|  
				p{1.38cm}|p{1.38cm}|p{1.38cm}| 
				p{1.38cm}|p{1.38cm}|p{1.38cm}| }
			\hline 
			\multicolumn{3}{|c|}{${N=20}$, ${M=10}$, and ${S=2}$} 
			&  \multicolumn{3}{c|}{${N=80}$, ${M=40}$, and ${S=8}$} 
			& \multicolumn{3}{c|}{${N=160}$, ${M=80}$, and ${S=16}$} \\
			\hline
			${R=1}$ & ${R=2.5}$ & ${R=4}$
			& ${R=1}$ & ${R=2.5}$ & ${R=4}$
			& ${R=1}$ & ${R=2.5}$ & ${R=4}$\\  
			\hline
			0.4\,/\,2.5\,/\,1.8 & 0.5\,/\,2.6\,/\,1.8 & 0.7\,/\,2.6\,/\,1.6 
			& 0.2\,/\,3.4\,/\,2.6 & 0.2\,/\,3.4\,/\,2.5 & 0.4\,/\,3.5\,/\,1.9 
			& 0.1\,/\,7.6\,/\,6.6 & 0.2\,/\,7.6\,/\,6.3 & 0.5\,/\,7.4\,/\,5.3  \\
			\hline
			0.4\,/\,57\,/\,38 & 0.5\,/\,58\,/\,36 & 0.7\,/\,58\,/\,26 
			& 0.2\,/\,53\,/\,40 & 0.2\,/\,54\,/\,39 & 0.4\,/\,54\,/\,27  
			& 0.1\,/\,38\,/\,33 & 0.2\,/\,37\,/\,31 & 0.5\,/\,37\,/\,26  \\
			\hline
		\end{tabular}\label{table:time}
	\end{center}
\end{table*}

To summarize the findings from the conducted experiments, the proposed VQ-based $\DeepQCS$ method obtains superior rate-distortion performance with orders of magnitude lower algorithm running time as compared to the conventional QCS methods, rendering $\DeepQCS$ a potential method for finite-rate communication of sparse signals with resource-limited encoding devices.

\section{Conclusion}\label{sec:conclusions}
We proposed the $\DeepQCS$ architecture, consisting of the encoder DNN, quantizer, and decoder DNN, for low-complexity acquisition of sparse sources through vector quantized noisy compressive measurements. A supervised SGD learning algorithm and techniques to overcome the non-differentiability of quantization were proposed for training the $\DeepQCS$ scheme. Simulation results showed the superior rate-distortion performance and algorithm complexity of the proposed $\DeepQCS$ scheme compared to standard QCS methods. These are desirable features to make $\DeepQCS$ as a potential candidate for rate-limited communication of sparse signals through QCS under limited encoding and decoding capabilities.

The present study opens several research avenues. First, it would be interesting to test the proposed method using real-world sparse signals to obtain insights about the performance in a practical scenario. As potential future work, different encoder/decoder DNN types could be considered. The system could be extended to incorporate lossy communication channels, calling for a design of a channel-aware QCS method to counteract the erroneous transmissions of the code words. An extension to a distributed QCS setting with multiple encoding devices has its relevance to model, e.g., a practical wireless sensor network.

\section*{Appendix}
The partial derivative of cost function $C$ in \eqref{eq:cost_training} with respect to level coefficient $v_{i}$, ${i=1,\ldots,I-1}$, is derived as
\begin{equation}\label{eq:grad_SHQ_level_appendix}
\begin{array}{ll}
\disp\frac{\partial{C}}{\partial{v_{i}}}\!\!\!\!\!&=\sum_{n=1}^{K}\sum_{n'=1}^{K}\disp\frac{\partial{C}}{\partial{{z}_{1,n'}}}\frac{\partial{{z}_{1,n'}}}{\partial{{a}_{J+1,n}}}\frac{\partial{{a}_{J+1,n}}}{\partial{v_{i}}}\\
\!\!\!\!\!&\overset{(a)}{=}\sum_{n=1}^{K}\sum_{n'=1}^{K}\disp\frac{\partial{C}}{\partial{{z}_{1,n'}}}\frac{\partial{{z}_{1,n'}}}{\partial{{a}_{J+1,n}}}\tanh\big(h{a}_{J,n}-s_{i}\big)\\
\!\!\!\!\!&\overset{(b)}{=}\sum_{n=1}^{K}\delta_{1,n}\tanh\big(h{a}_{J,n}-s_{i}\big),\\
\end{array}
\end{equation}
where equality $(a)$ follows from the differentiation of the SHQ function with respect to $v_{i}$ as $\frac{\partial{{a}_{J+1,n}}}{\partial{v_{i}}}=\frac{\partial{\gamma_{J+1}({a}_{J,n})}}{\partial{v_{i}}}
=\frac{\partial{\big(\sum_{i=1}^{I-1}v_{i}\tanh\big(h{a}_{J,n}-s_{i}\big)\big)}}{\partial{v_{i}}}
=\tanh\big(h{a}_{J,n}-s_{i}\big)$; equality $(b)$ follows i) because by ${\zb_{1}=\ab_{J+1}}$, we have ${\frac{\partial{{z}_{1,n'}}}{\partial{{a}_{J+1,n}}}=\frac{\partial{{a}_{J+1,n'}}}{\partial{{a}_{J+1,n}}}=1}$, if ${n=n'}$, and otherwise $0$, and ii) from substitution ${\delta_{1,n}=\frac{\partial{C}}{\partial{{z}_{1,n}}}}$ from \eqref{eq:grad_dec}.

The partial derivative of $C$ with respect to shift coefficient  
$s_{i}$, ${i=1,\ldots,I-1}$, is derived as
\begin{equation}\label{eq:grad_SHQ_shift_appendix}
\begin{array}{ll}
\disp\frac{\partial{C}}{\partial{s_{i}}}\!\!\!\!\!&=\sum_{n=1}^{K}\sum_{n'=1}^{K}\disp\frac{\partial{C}}{\partial{{z}_{1,n'}}}\frac{\partial{{z}_{1,n'}}}{\partial{{a}_{J+1,n}}}\frac{\partial{{a}_{J+1,n}}}{\partial{s_{i}}}\\
\!\!\!\!\!&\overset{(a)}{=}\sum_{n=1}^{K}\sum_{n'=1}^{K}\disp\frac{\partial{C}}{\partial{{z}_{1,n'}}}\frac{\partial{{z}_{1,n'}}}{\partial{{a}_{J+1,n}}}v_{i}\tanh'_{s_{i}}\big(h{a}_{J,n}-s_{i}\big)\\
\!\!\!\!\!&\overset{(b)}{=}\sum_{n=1}^{K}\delta_{1,n}v_{i}\tanh'_{s_{i}}\big(h{a}_{J,n}-s_{i}\big),
\end{array}
\end{equation}
where equality $(a)$ follows from $\frac{\partial{{a}_{J+1,n}}}{\partial{s_{i}}}
=\frac{\partial{\big(\sum_{i=1}^{I-1}v_{i}\tanh\big(h{a}_{J,n}-s_{i}\big)\big)}}{\partial{s_{i}}}
={{v_{i}}}\tanh'_{s_{i}}\big(h{a}_{J,n}-s_{i}\big)$,
where $\tanh'_{s_{i}}(\cdot)$ denotes the derivative of $\tanh(h{a}_{J,n}-s_{i})$ with respect to $s_i$, given as $-4/{\big(\exp\{h{a}_{J,n}-s_{i}\}+\exp\{-h{a}_{J,n}+s_{i}\}\big)^2}$; equality $(b)$ follows similarly as step $(b)$ in \eqref{eq:grad_SHQ_level_appendix}.

\ifCLASSOPTIONcaptionsoff
\newpage
\fi

\bibliographystyle{IEEEtran}
\bibliography{main}

\end{document}